\documentclass[11pt,letterpaper]{article}

\usepackage{fullpage} \usepackage{booktabs} 
\usepackage{pslatex} \usepackage{hyperref} \usepackage{url}
\usepackage{apacite} \usepackage{amsmath} \usepackage{subcaption}
\usepackage[utf8]{inputenc} \usepackage{pgfplots} \pgfplotsset{compat=newest}
\usepgfplotslibrary{groupplots} \usepackage{wrapfig} \usepackage{bigfoot}
\usepackage[export]{adjustbox}  \usepackage{authblk}
\usepackage{float}

\usepackage{graphicx}

\usepackage{gb4e}  
\noautomath

\title{
Paths to Polarization: \\
How Extreme Views, Miscommunication, and Random Chance Drive Opinion Dynamics}

\author[1,*]{Matthew~A.~Turner} \author[1]{Paul~E.~Smaldino}

\affil[1]{\footnotesize Cognitive and Information Sciences, University of
California, Merced}
\affil[*]{\footnotesize Corresponding author. Email: mturner8@ucmerced.edu}
\date{}

\begin{document} \maketitle

\begin{abstract} 
\noindent Understanding the social conditions that tend to increase or decrease polarization is
important for many reasons. 
We study a network-structured agent-based model of opinion dynamics, extending a model previously introduced by Flache and Macy (2011), who found that polarization appeared to increased with the introduction of long-range ties but decrease with the number of salient opinions, which they called the population's ``cultural complexity.'' 
We find the following. 
First, polarization is strongly path dependent and sensitive to stochastic variation. 
Second, polarization depends strongly on the initial
distribution of opinions in the population. In the absence of extremists,
polarization may be mitigated. 
Third, noisy communication can drive a population toward more extreme opinions and even cause acute polarization. 
Finally, the apparent reduction in polarization under increased ``cultural complexity'' arises via a particular property of the polarization measurement, under which a population containing a wider diversity of extreme views is deemed less polarized. 
This work has implications for understanding the population dynamics of beliefs, opinions, and polarization, as well as broader implications for the analysis of agent-based models of social phenomena. 
\end{abstract}

\small{\textbf{Keywords:} polarization; opinion dynamics; small-world networks; cultural complexity; agent-based models}
\section{Introduction}

Diversity of opinions in a community is often difficult to maintain. Iterative exposure, norm
enforcement, and psychological biases for conformity can drive consensus within
a group \cite{degroot1974reaching,deffuant2000mixing,henrich1998evolution,smaldino2015social,efferson2008conformists,muthukrishna2016and}.
On the other hand, in-group bias, outgroup aversion, and the
tendency to further differentiate ourselves from those deemed different may lead to the
emergence of strong inter-group differences \cite{tajfel1971social,Lord1979,carley1990group,Axelrod1997,mark1998beyond,mcelreath2003shared,Dandekar2013,gray2014emergence,smaldino2017adoption}.
Such differences
can lead to polarization in opinions under certain conditions.  Understanding
the social conditions that tend to increase or decrease polarization is
important for many reasons. Primary among these is that a functioning 
democratic society
depends on clear communication among the citizenry, which is impeded by the
mismatch in norms, the differential interpretation of facts, and the
dehumanization that polarization can engender (see \citeA{PewResearchCenter2017}
for a current analysis of these dynamics in the United States). 
The maintenance of social differences in the form of cliques and clubs 
may be inevitable, but cooperation depends on transcending differences. 

We take a network theoretic approach to studying the conditions for polarization in an agent-based model of opinion dynamics. Empirical research on the population dynamics of opinions is challenging and must be supplemented by formal modeling 
\cite{flache2017models}. 
Models reduce complex systems to ones that are tractable using mathematical or computational analysis, and allow for the exploration of replicate and counterfactual scenarios. Of course, the conclusions we draw from our models depend essentially on the assumptions of those models, and so caution must be taken when using model results to make inferences about empirical phenomenon. For example, \citeA{smaldino2012human} analyzed models of human mate choice and showed that very different individual decision rules could be fit to almost any empirical outcome by modulating assumptions about the population structure that had been ignored in prior analyses. When considering an important phenomena such as polarization, similar caution must be exercised, as we will demonstrate. 

Our analysis extends the work of \citeA{Flache2011}, who used a
network-structured model of opinions and biased influence (hereafter
the FM model) to study polarization. Network ties in this model exist between individuals as an indicator of social influence. Like several other models of opinions and beliefs, they
operationalized the well-known phenomena of {\em biased assimilation} \cite{Lord1979,Dandekar2013}, the
tendency for an individual to become more similar to those to whom they are
similar, and to become more distinct from those with whom they already differ.
Some empirical studies support the assumption of both positive and negative biased assimilation
\cite[e.g.]{Adams2005, Hart2012}. Other empirical studies failed to find evidence of negative
biased assimilation at work where computational studies suggested it would be 
\cite[e.g.]{Takacs2016, Boxell2017a}. 
Of course, if further empirical research turns out to invalidate that assumption, then our model conclusions must also be re-examined, as with any theoretical model \cite{Smaldino2017}.
Flache and Macy found that, when compared with a highly clustered population structure, the addition of
long-range ties could dramatically increase polarization. 
When individuals were clustered into relatively isolated groups, they tended to converge to local consensus while maintaining diversity in the population at large. However, the addition of long-range ties increased exposure to substantially different opinions. 
Whether by attractive or repulsive forces, these long-range ties tended to
drive opinions more toward their extreme values, resulting in increased
polarization.  Another important result was that the extent of ``cultural
complexity''---the number of orthogonal traits that are important to individuals
in assessing their similarities and differences with others---mitigated
polarization. When the number of traits was large, polarization was
reduced. \citeA{DellaPosta2015} used a variant of the FM model to explain data from the General Social Survey indicating that arbitrary traits tend to become associated with polarized identity groups, leading to
often-puzzling stereotypes such as ``latte-drinking liberals" and ``bird-hunting conservatives."  

If we take the results of \citeA{Flache2011} at face value, two possible recommendations for the reduction of polarization readily emerge.  First, we might try to reduce the
number of long-range ties in our social network. This is made difficult 
due to the pervasive
influence of internet social media~\cite{PewResearchCenter2016,PewResearchCenter2018}. 
Second, we might attempt to broaden the
number of domains in the public discussion, so that points of agreement are
easier to discover. This is also challenging, due to the increasingly fractured
media landscape in which niche interests are increasing and common knowledge
diminishing \cite{Pew2014}. 
However, challenging is not the same thing as impossible. We must ask, then: 
How seriously should we take these recommendations? Might there be
other solutions available?

To address these questions we perform new analyses of the FM model and reveal several additional factors
influencing polarization. First, polarization is almost always a probabilistic
occurrence. Even when parameter exploration appears to reveal regularities in
polarization, specific outcomes are strongly path dependent. Indeed, there is
often a wide range of possible outcomes even given identically repeatable
starting conditions, due to stochasticity in the dynamics of interactions. This
result highlights potential limits of our ability to make reliable
predictions about polarization in any particular social system. Complex systems are
often stochastic, and something that increases or decreases average polarization
in a simulation is not guaranteed to do so in reality.
Second, resultant polarization depends strongly on the initial
distribution of opinions in the population. In the absence of extremists,
polarization may be mitigated. This highlights the well-known danger of
extremists and suggests new routes to avoiding polarization. More broadly, we
show that too much diversity
of extreme opinions makes polarization more likely.
Third, noisy communication can drive a population toward more extreme opinions and even cause acute polarization. Cooperation and consensus-building depend on individuals finding common
ground, which can be jeopardized even in the presence of unbiased error \cite{Clark1996}. 
Finally, we show that the apparent reduction in polarization under increased ``cultural complexity'' arises via a particular property of the polarization measurement, 
under which a population containing a wider diversity of extreme views is 
deemed less polarized. Although this may often be a reasonable assumption, 
it highlights the need for caution in our measurement of complex social phenomena.

\section{Model}

\subsection{Modeling individuals and their opinions} 

Our model is an extension
of one presented by \citeA{Flache2011}, and shares many general features with
other models of opinion dynamics in structured populations \cite{Nowak1990,carley1990group,Axelrod1997,mark1998beyond,mark2003culture,Dandekar2013,DellaPosta2015,battiston2017layered}.  The population is modeled as a network of individuals (or agents), each of whom is defined by a vector of opinions. The
size of this vector, $K$, is called the ``cultural complexity,'' 
and may be more descriptively explained as the number of  opinions that are
important to individuals in assessing their similarities and differences with
others. Opinions can present political views, religious or moral values,
artistic tastes, or myriad other beliefs. The opinion of agent $i$ on issue $k$
$(1 \leq k \leq K)$, $s_{ik}$, is operationalized as a real number implicitly bounded in
$[-1, 1]$ by smoothing (Equation \ref{eq:smoothed-update}). 
In Flache and Macy's original analysis, all opinions were initialized as random
draws from the uniform distribution $U(-1,1)$. In order to study the importance
of initially extreme opinions, each initial opinion is here drawn instead from
$U(-S, S)$, where $0 < S \leq 1$.   

\subsection{Modeling social influence} 

The aggregation of the $K$ opinions held by an agent 
determines its coordinates in opinion space.
We adopt the FM model's measure of distance between agents $i$ and $j$,

\begin{equation}
  d_{ij} = \frac{1}{K}\sum_{k=1}^{K} |s_{jk,t} - s_{ik,t}|.
\end{equation}
\noindent
Distance thus defined measures the average absolute difference across 
opinion coordinates.  Agents are nodes in a network, 
with an edge between agents reflecting a
relationship and an opportunity for the agents to influence one another. The
magnitude and direction of that influence is characterized by the {\em weight}
of each edge. Weights are determined by the relative opinions of the
two agents, as measured by their distance, and so can change dynamically. Positive weights represent positive
influence, in which agents become closer in their opinions, while negative
weights represent the tendency toward differentiation. For descriptive convenience, if two agents are
connected with a positive weight, they could be considered ``friends'' and if
the weight is negative they could be considered ``enemies.'' In reality, no assumptions about such clear social roles are necessary. The weight of an edge
between agents $i$ and $j$ is given by 

\begin{equation} 
  w_{ij,t+1} = 1 - d_{ij,t}.  
\end{equation} 
\noindent
So, if the opinions of agents $i$ and $j$ are separated by $d_{ij} < 1$, 
the agents are friends and will harmonize their opinions. If $d_{ij} > 1$, 
the agents are enemies, and will drive each other's opinions to more extreme levels.
This weighting rule embodies the psychological phenomena of {\em biased assimilation},
in which similar individuals grow more similar and dissimilar individuals grow
further apart after interacting \cite{Lord1979}. This is a common assumption in models
of social influence \cite{Hegselmann2002,Flache2011,Dandekar2013}). 
It should be noted that while the
empirical evidence for biased assimilation is quite strong, and spans almost four decades, it is less clear how coherence on various opinions or beliefs affects influence on orthogonal opinions or beliefs. The assumption in this model is that it is only average distance in opinions that matters.  

At time $t+1$, agents update their opinions by adding the average 
influence from all neighbor agents.  For each opinion $k$, 
agent $i$ uses the following update rule: 

\begin{equation} 
  s_{ik,t+1} = s_{ik,t} + \Delta s_{ik,t} \left(1 - \text{sgn}(s_{ik,t}) s_{ik,t} \right), 
  \label{eq:smoothed-update}
\end{equation} 
\noindent
where

\begin{equation}
  \Delta s_{ik,t} = \frac{1}{2N_i} \sum_{j \neq i} w_{ij,t} (s_{jk,t} - s_{ik,t}) + \epsilon.  
\end{equation} 
\noindent
Here, $N_i$ is the number of agents
with which agent $i$ shares an edge, and $\epsilon$ is a noise term that
reflects errors in the communication of opinions. This term is in each
instance drawn at random from a normal distribution with a mean of zero and a
standard deviation of $\sigma$. We conceptualize updating to be the result of
agents sensing the communicated opinions of neighbors. Furthermore, we 
conceptualize this $\sigma$ as representing noise either in an agent 
sensing the opinions of other agents, noise in agents communicating their
opinions, or both. In their original study \citeA{Flache2011} considered only scenarios without noise ($\sigma = 0$).  
Time in the model progressed in discrete time steps. At each time step, each agent's opinions were updated asynchronously in random order to avoid well-known artefacts that often accompany simultaneous agent updating. 

\begin{figure}[H]
  \centering
  \begin{subfigure}[t]{\textwidth}
    \centering
    \includegraphics[width=.5\textwidth]{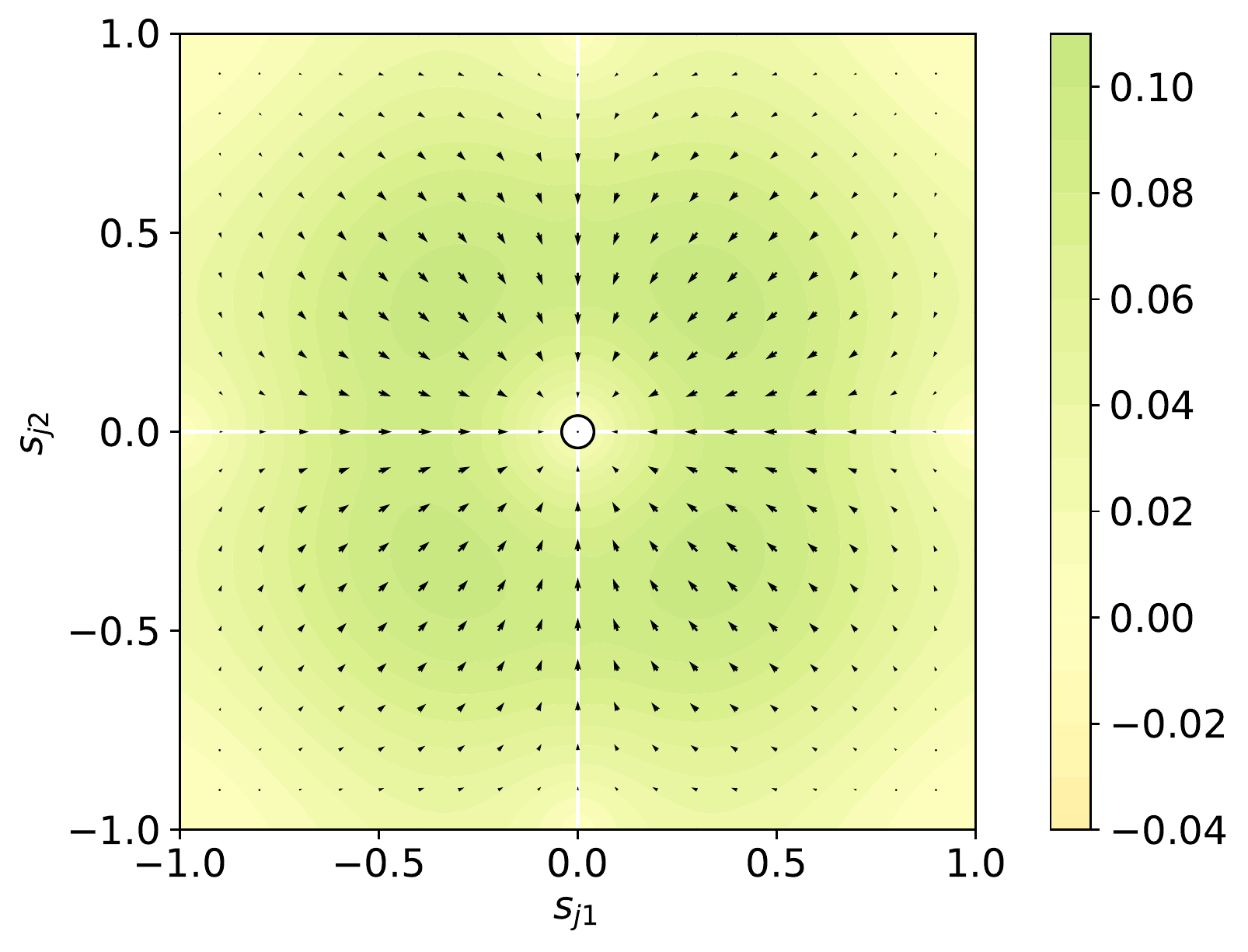}
    \caption{Influence of agent at origin.}
  \end{subfigure}\\[2em]
  \begin{subfigure}[t]{\textwidth}
    \centering
    \includegraphics[width=.5\textwidth]{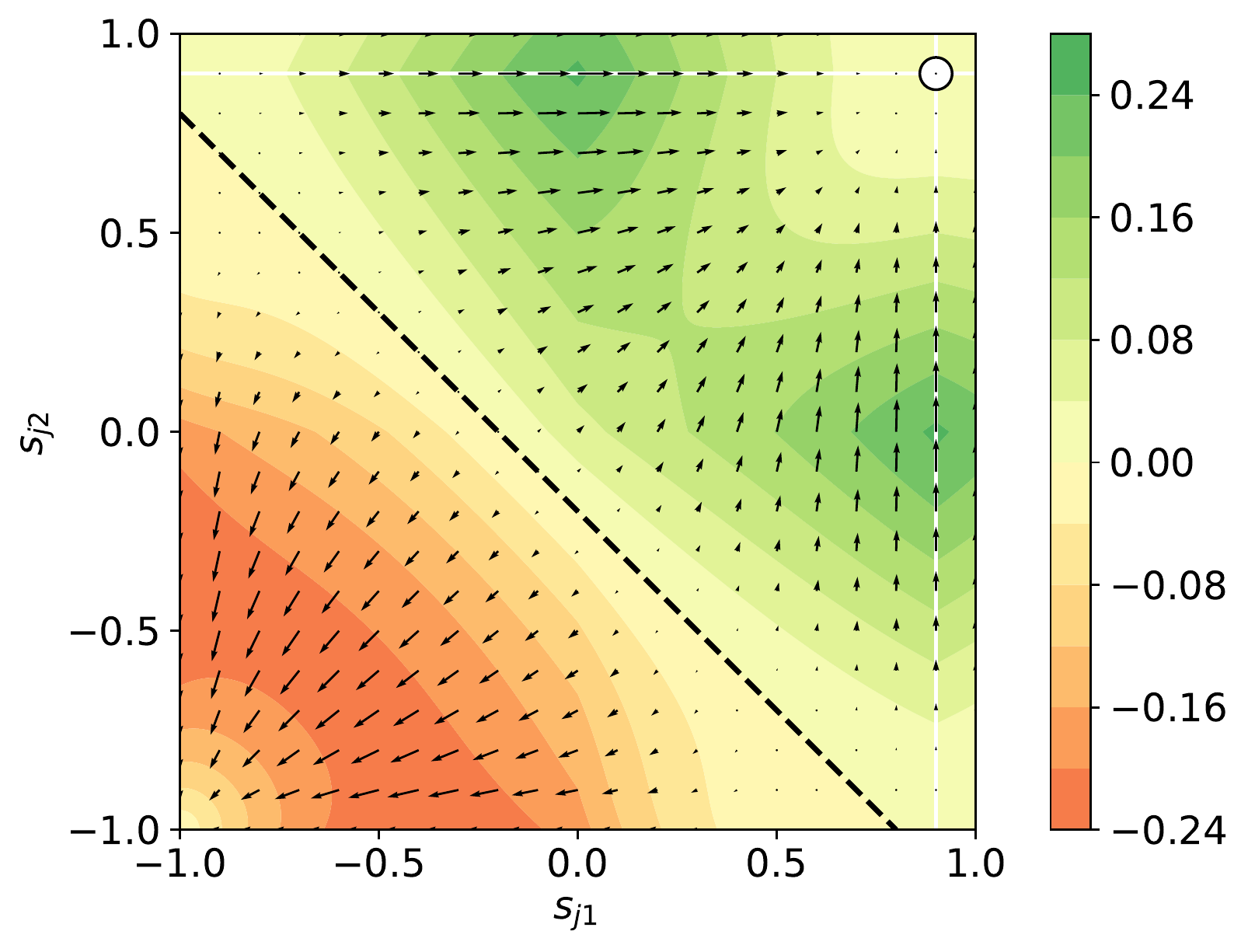}
    \caption{Influence of agent at (0.9, 0.9).}
  \end{subfigure}
  \caption{Influence by one agent on another changes 
    depending on the location of each agent. This illustrates the influence
    exerted by a central agent (white circle) on another agent at different
    locations in opinion space. 
  }
  \label{fig:dyad_influence}
\end{figure}

It is worth noting a few immediate consequences of these update equations.  
First, agents with extreme opinions in 
dimension $k$ will 
tend to make smaller changes to those opinions
because of the smoothing factor $(1 - \text{sgn}(s_{ik,t})s_{ik,t})$. In other words, extreme opinions will be harder to change. 
Second, there are two opposing factors that modulate the magnitude of 
influence between two agents. On the one hand, edge weight is maximal when agents' opinions are very similar. 
On the other hand, $\Delta s_{ik,t}$ (which Flache and Macy refer to as the ``raw" state change) increases the more agents' opinions differ, presumably because larger distances provide larger room for change, with a mathematical form drawn from psychological models of reinforcement learning \cite{rescorlaw72,sutton1998reinforcement}. Influence will therefore be maximal for agents who are an intermediate distance apart in opinion space. 
To facilitate an intuitive understanding of dyadic interactions, we illustrate the strength of influence on
agent opinions in $K=2$ opinion space 
in Figure~\ref{fig:dyad_influence}. 
We see that an agent with opinions at the origin of opinion space has only 
a moderate, attractive influence on other agent opinions in the opinion space.
Agents at the corners of opinion space are barely influenced by a central 
opinion vector. When we consider the influence of an agent opinion nearer
to the corner, at $\vec s_i = (0.9, 0.9)$, we see that there is a clear
line where relationships switch from friend to enemy ($s_{j2} = s_{j1} - 0.2$). 
Due to the co-mingling of effects described above, 
there is a varied and non-monotonic landscape of influence.

\subsection{Measuring Polarization}

There are a multitude of measures for polarization \cite{Bramson2016} and no single measure is widely agreed upon. We follow \citeA{Flache2011} and define polarization at 
time $t$ to be the variance of all distances between agents,

\begin{equation}
  P_t = \text{var}(d_{ij,t})
\end{equation}
\noindent
This metric has the advantage of simple interpretation. If half of all agents are
in one corner of opinion space and the other half of agents are in the
opposite corner, then the population is maximally polarized. As agent opinions
spread to other corners and to other regions of opinion space, polarization
will decrease. One disadvantage is that more general patterns of clustering,
as would be detected using various machine learning clustering algorithms, will go undetected. In the final subsection of our Results, we illustrate another limitation of this metric. Nonetheless, we generally find that it is a useful and suitable operationalization for the concept of polarization.

\subsection{Network structure}

Our network structures are taken from Flache and
Macy's (2011) Experiment 2. We begin with the connected caveman network
structure introduced by \citeA{Watts1999}. Specifically, we consider a network of
$N = 100$ agents, grouped into 20 fully connected clusters (caves) of five
agents each. These caves are arranged on a circle, and for each cave one edge
is selected at random and rewired to connect to a random agent in the cave
immediately to the right of the focal cave. This network has the appearance of tight-knit communities with weak ties to neighboring communities. The connected caveman network is highly clustered, meaning that if two agents are both neighbors of another single agent, there is a high probability that those two agents are also neighbors. However, relative path length is considerably greater in a connected caveman graph
than for a totally random graph. 

\begin{figure}[H]
  \centering
  \begin{subfigure}[t]{0.45\textwidth}
    \centering
    \includegraphics[width=\textwidth]{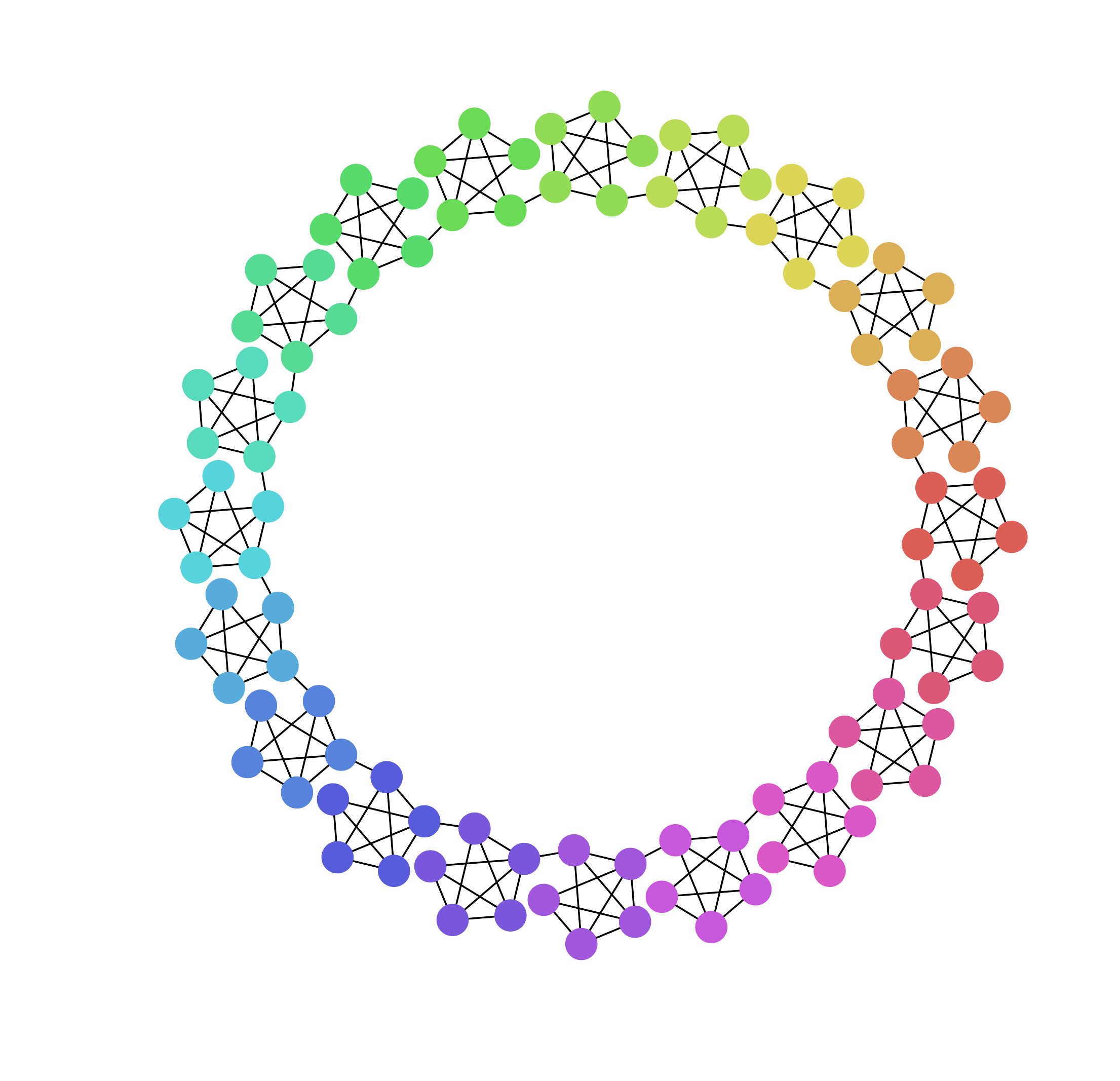}
    \caption{Connected caveman graph before long-range ties added.}
  \end{subfigure}
  \begin{subfigure}[t]{0.45\textwidth}
    \centering
    \includegraphics[width=\textwidth]{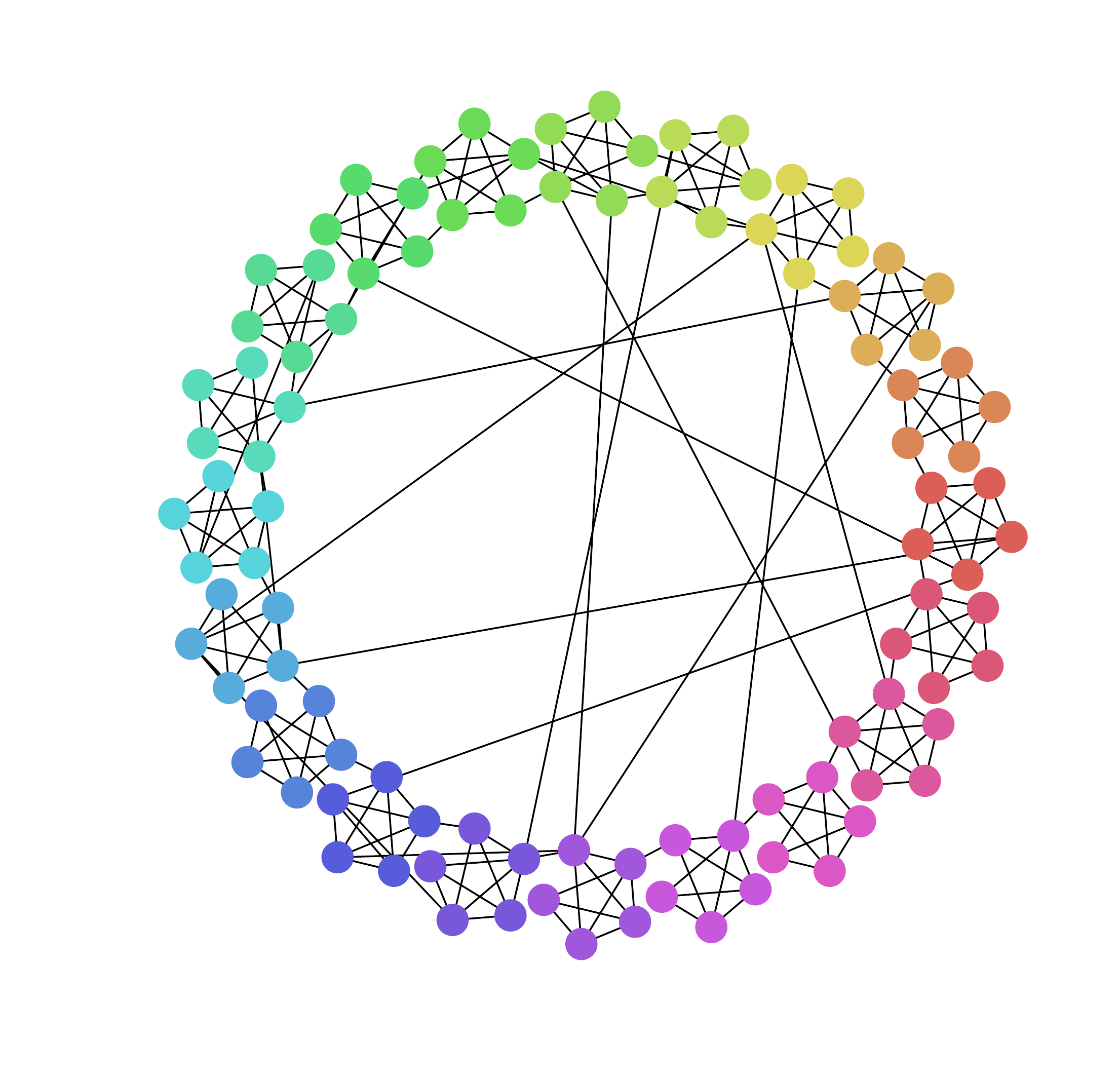}
    \caption{After long-range ties added.}
  \end{subfigure}
  \caption{Connected caveman network with and without twenty long-range ties.
    Colors represent cave membership.
  }
  \label{fig:network}
\end{figure}

To assess the influence of adding long-range ties, we then
consider a network for which 20 additional edges are added between randomly
selected pairs of agents from across the entire network (Figure~\ref{fig:network}). 
Long-range ties are added at $t = 2000$ to give the local communities (caves) 
time to yield enclaves of conformity that differ slightly from their neighboring enclaves,
following \citeA{Flache2011}. The long range ties reduce the average path length of the network while retaining high clustering, yielding networks with ``small-world'' properties \cite{Watts1999}.

Finally, as a way to control for the effect of simply adding additional ties, we also consider the connected caveman network with {\em short-range} ties. In this case a randomly selected agent from each cave (who is not already connected to another cave) is
connected to a random agent in the cave immediately to the right of the focal cave. Unless stated otherwise, all of our analyses were restricted to the connected caveman network with long-range ties, as this was the network structure found by \citeA{Flache2011} to maximize polarization. 

\subsection{Computational experiments} 
Below we present the results of our computational experiments. For all parameter combinations we ran 100 simulations of the model, with data collected after $10^4$ time steps. This was always sufficient time for the system to settle down into a relatively stable pattern (true equilibria were not always reached due to the stochasticity inherent in the model). By calculating the difference in polarization on the final timestep for all simulations and finding all to be sufficiently small, we confirmed that 10$^4$ timesteps was sufficient to achieve stable behavior across all simulations. 
We first replicate the major result of \citeA{Flache2011} that polarization increases with the addition of long-range ties but decreases with increasing cultural complexity, $K$. We then perform three sets of experiments: 
\begin{enumerate}
\item {\em Quantifying variation.} We take a closer look at the variation among simulation runs, and explore path dependence on the road to polarization. 
\item {\em Reducing extremism.} We investigate values of $S < 1$, in which the initial distribution of opinions is less extreme.  
\item {\em Adding noise.} We investigate values of $\sigma > 0$, in which communication about opinions is noisy and influence is therefore more stochastic.  
\end{enumerate}
Unless stated otherwise, all simulations used a connected caveman network with random long-range ties, $S = 1$, and $\sigma = 0$. 
Model and analysis code is available on GitHub at \url{https://github.com/mt-digital/polarization}.

\section{Results}

\begin{figure}[H]
  \centering
    \includegraphics[width=0.6\textwidth]{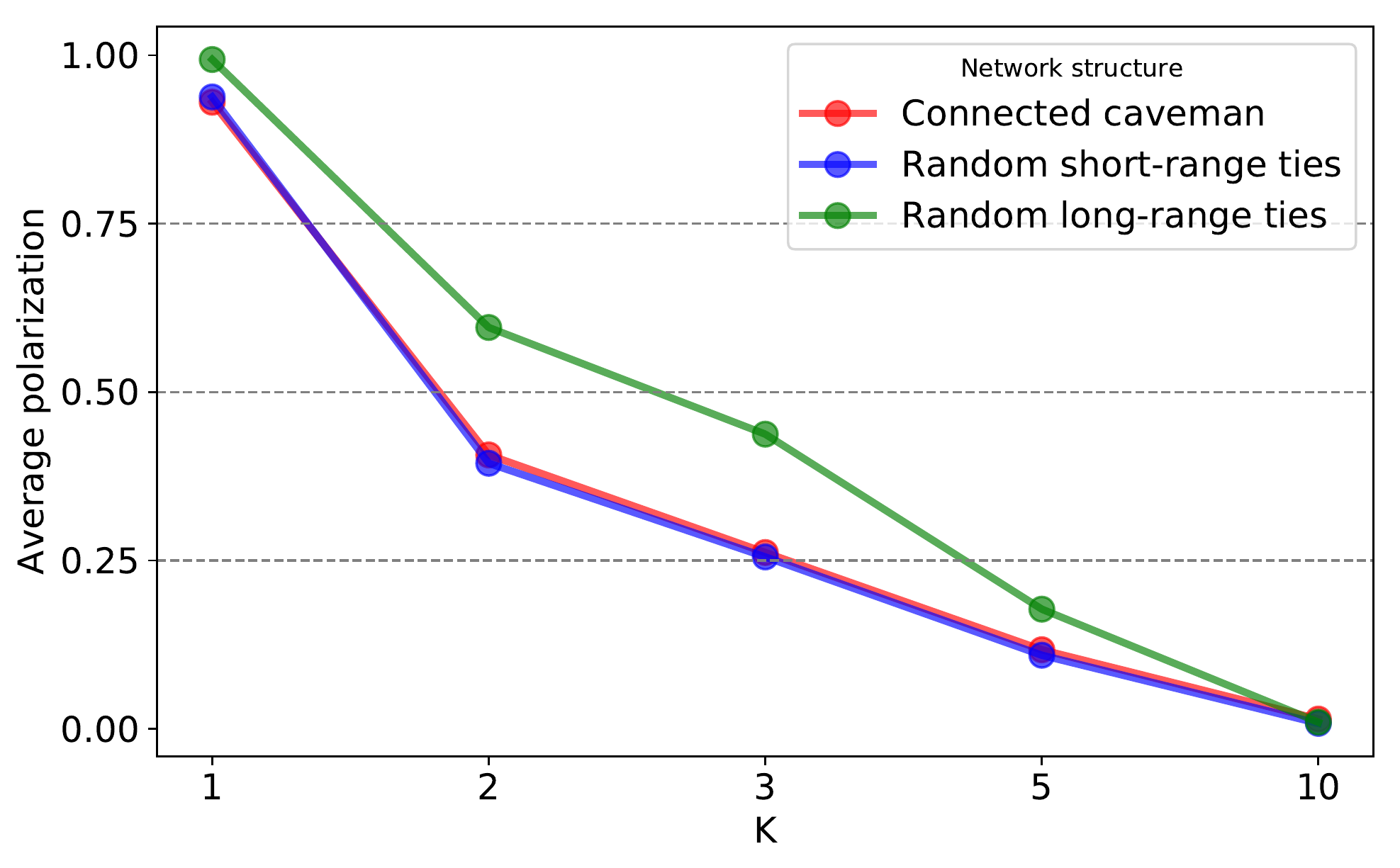}
  \caption{Reproduction of Figure 12b of Flache and Macy (2011). Average
    polarization decreases with $K$. However, as shown in subsequent figures,
    this does not mean trials with high polarization never obtain for large $K$. Average
    taken over 100 trials.
  }
  \label{fig:p_vs_K_fm}
\end{figure}

In their original analysis of the FM model, \citeA{Flache2011} found two main causes of polarization. First, random long-range ties decreased the average path length of the network and increased the average polarization of the system across trials. Second, 
average polarization across trials decreased with increasing cultural complexity, $K$.
We replicated these results, as illustrated in Figure~\ref{fig:p_vs_K_fm}. 
The remainder of this section is dedicated to novel results. The first three subsections show results of new analyses of the original FM model. The final subsection shows our analysis of the FM model modified to include communication noise.

\subsection{Polarization is probabilistic and path-dependent}

\begin{figure}[H]
  \centering
    \begin{subfigure}[t]{\textwidth}
      \centering
      \includegraphics[width=.7\textwidth]{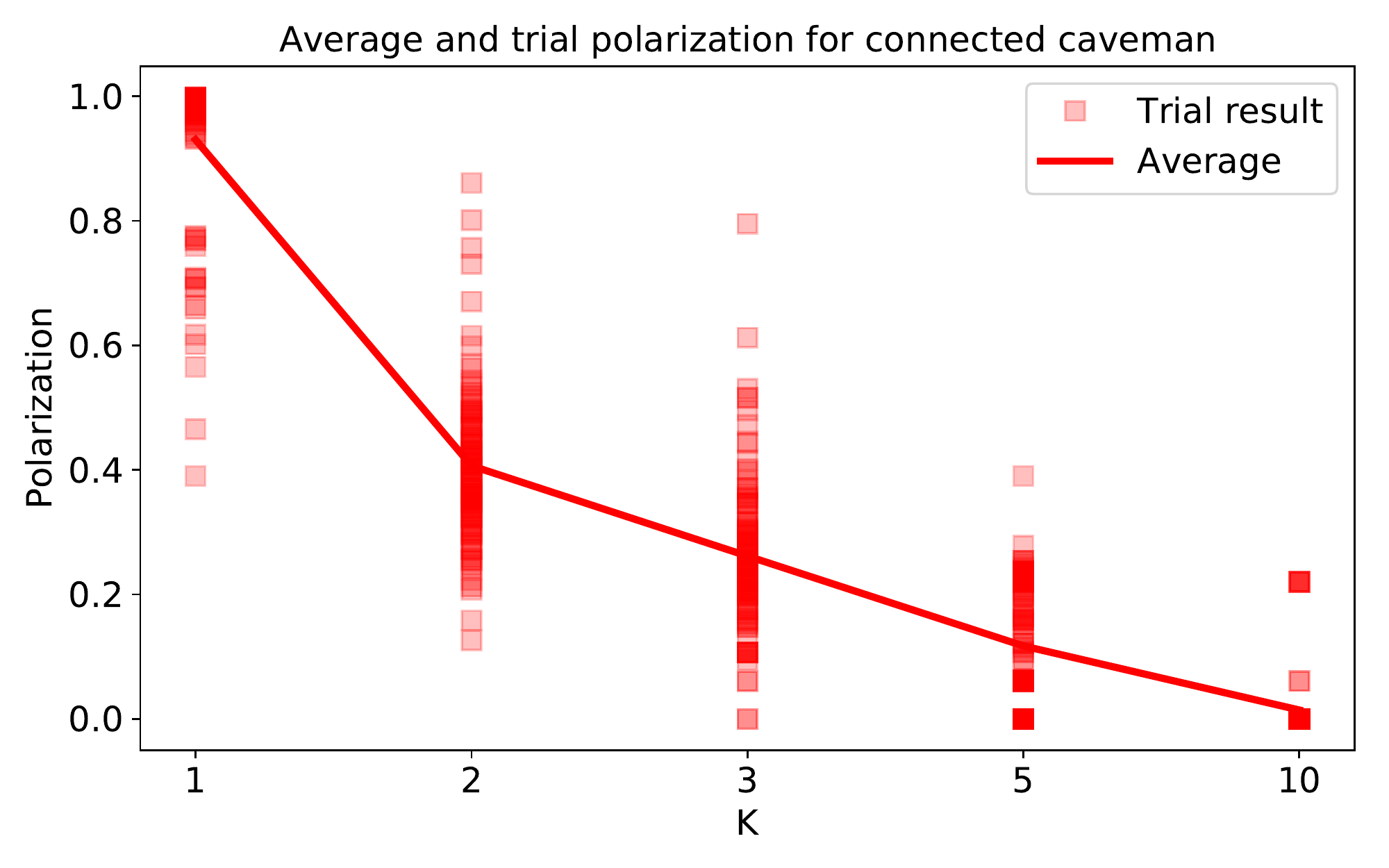}
      \caption{Non-random connected caveman network.}
      \label{fig:connected-caveman-trials}
    \end{subfigure}
    \begin{subfigure}[t]{\textwidth}
      \centering
      \includegraphics[width=.7\textwidth]{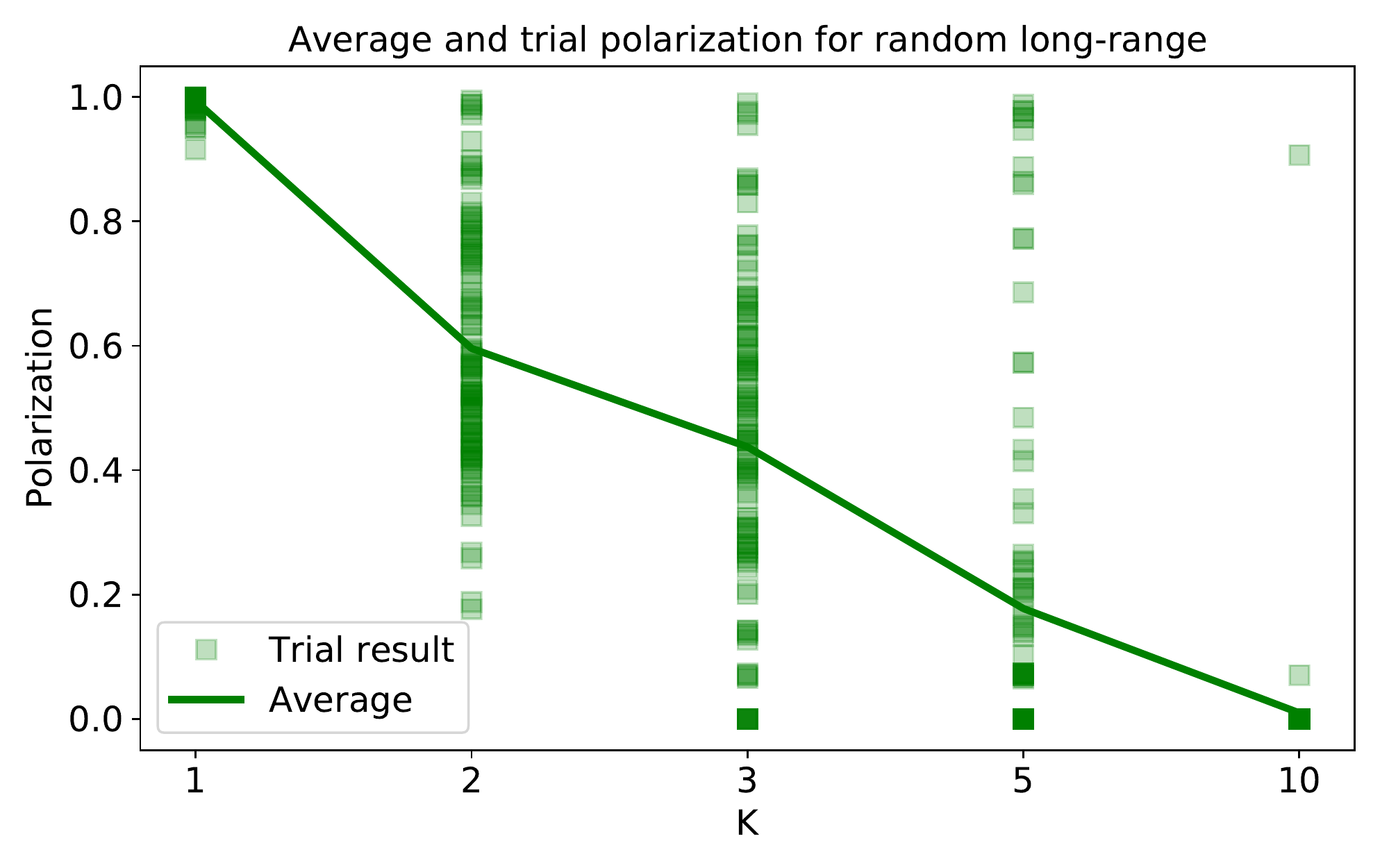}
      \caption{Randomized connected caveman network with long-range random ties added at iteration 2000.}
      \label{fig:random-anyrange-trials}
    \end{subfigure}
  \caption{Results of individual model runs under different network conditions. 
    The averages of these were shown in Figure~\ref{fig:p_vs_K_fm}.
    Even in the non-random connected caveman structure, there is 
    variation in the final polarization for different values of $K$. Highly
    polarized final states may obtain even for large $K$. 100 trials are shown for each 
    network condition. Solid lines indicate the average across all trials.}
  \label{fig:single-experiments-over-k}
\end{figure}

Averages do not carry information about variation between trials. Here we explore that variation. Figure~\ref{fig:single-experiments-over-k} shows the polarization for each of the individual trials averaged in Figure~\ref{fig:p_vs_K_fm}. We see a lot of variation around those averages, and that although polarization was low in all cases for large $K$, there are still individual trials for which polarization was high across all three network structures. 

\begin{figure}[H]
  \centering
    \includegraphics[width=.5\textwidth]{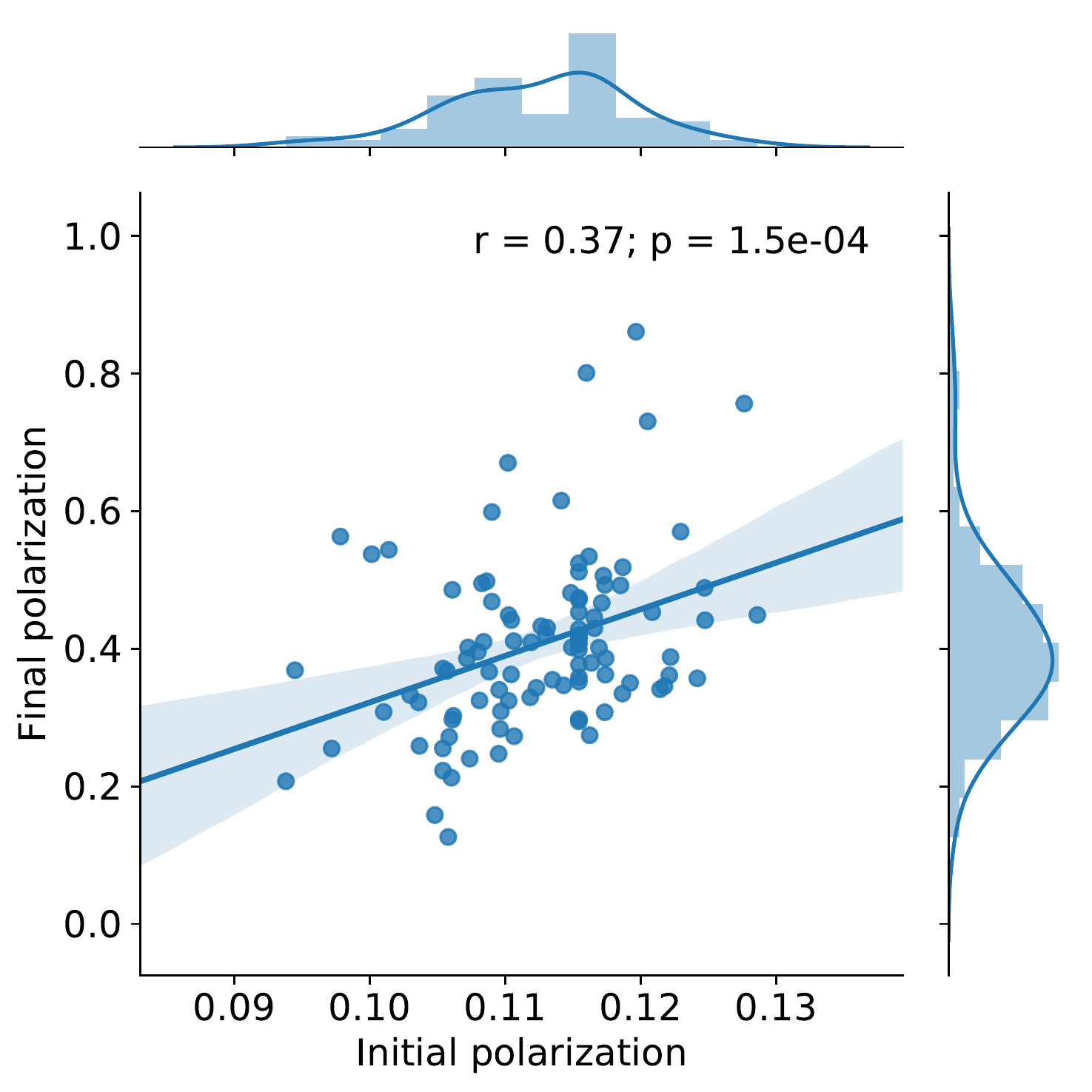}
  \caption{Regression of final polarization against initial polarization 
    for $K=2$ in the non-random connected caveman network configuration.
    Final polarizations are same as in the $K=2$ column of 
    Figure~\ref{fig:connected-caveman-trials}. 100 trials are shown. The
    top histogram shows the distribution of initial polarization across
    trials. The right histogram shows the distribution of final polarization
    across trials.}
  \label{fig:final-initial-pol-regplot}
\end{figure}

\begin{figure}[H]
  \centering
    \includegraphics[width=.75\textwidth]{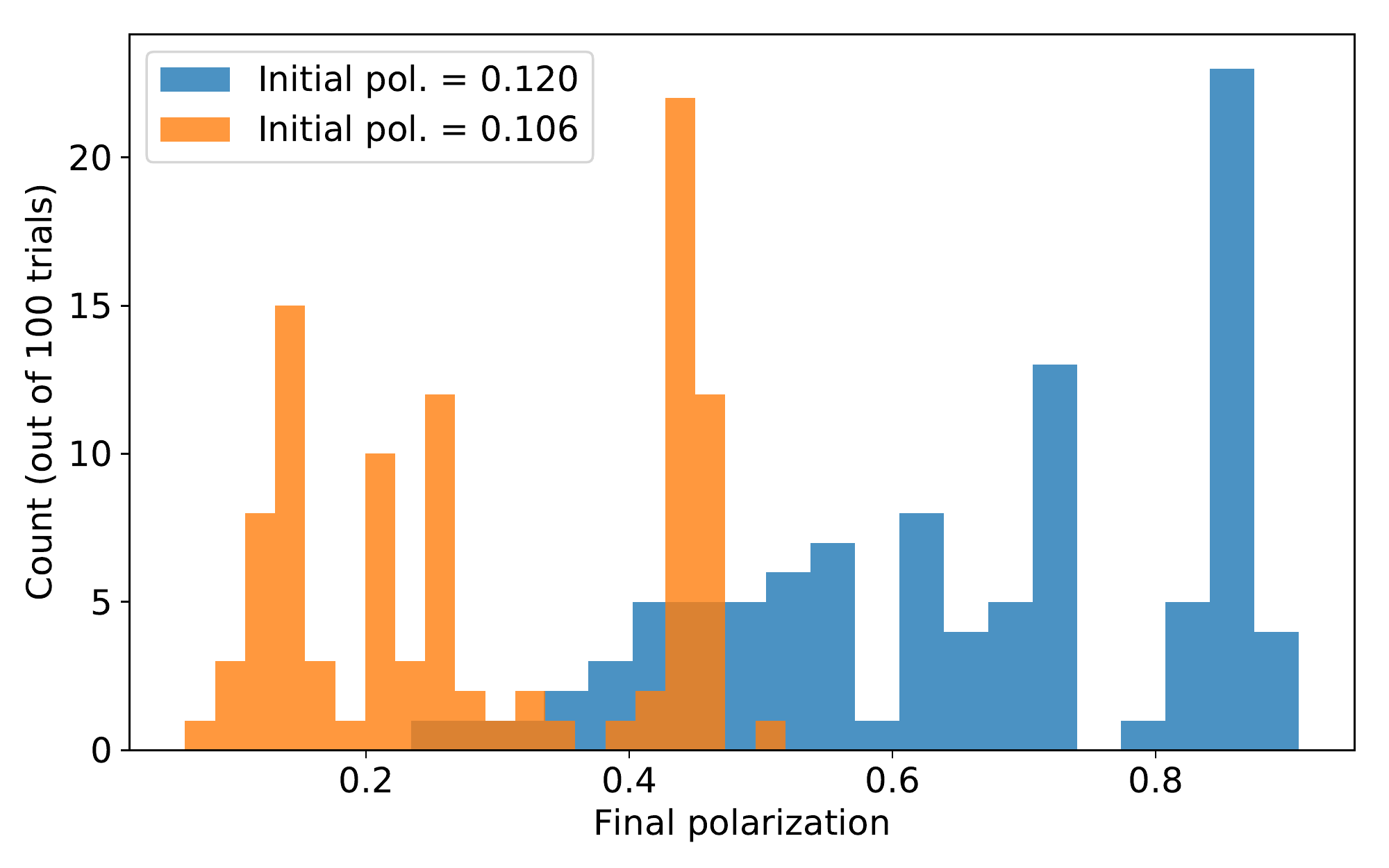}
  \caption{Distribution of final polarizations at $t = 10^4$
    starting from initial conditions of either maximum or minimum polarization taken from the 
    the connected caveman trials with $K = 2$.
  }
  \label{fig:highpol-histogram}
\end{figure}

In addition to the demonstrated influence of the overall network structure, three possible sources of  variation in system polarization are (1) the initial distribution of agent opinions, (2) the initial distribution of how agent opinions are clustered on the network, and (3) the update path---the order in which weights or agent opinions are updated. We performed additional analyses to investigate the contributions from each of these three factors, focusing on the initial distribution of agent opinions. We studied the non-random connected caveman network so as to keep network structure constant across trials, and for simplicity we restricted this analysis to $K=2$. 
Due to the nature of our polarization measure, at initialization the system will have some non-zero degree of polarization, which will vary depending on the random draws of agents' initial opinions. Over 100 trials, we compare the initial polarization of the system to the final polarization.    
We found a significant, if relatively small, correlation between the initial and final polarization of agent opinions, $r^2=.137$ (Figure~\ref{fig:final-initial-pol-regplot}). 
This means that the level of initial polarization accounts for only about 14\% of the 
variation in final polarizations.
It seems, then, that initial clustering of agent opinions and the stochasticity of the update path account for a large portion of the variability. In order to delineate the contributions of these two remaining factors to the overall variability in polarization, we considered the previously discussed simulations and ran 100 replicate trials with the initial conditions taken from the trials with the lowest and highest initial polarization. In other words, for each of two conditions, we ran replicate simulations with the exact same starting conditions between trials. Any variation in outcomes must therefore be due to stochasticity in the update paths. 
For example, if two opposing extremists influence a disjoint set of moderates disproportionately 
often, polarization will increase. The results are shown in Figure~\ref{fig:highpol-histogram}.  
Final polarization was clearly biased by the initial polarization (average final polarization across trials  
was 0.66 for the larger initial polarization, and 0.290 for the smaller initial polarization), but showed considerable variability. 
In other words, a large proportion of the variation between trials was due to stochasticity not in the initial configuration of the population, but to stochasticity in the transient dynamics of agent interactions.

\subsection{The absence of initially extreme opinions reduces polarization}

\begin{figure}[H]
  \centering
    \includegraphics[width=0.65\textwidth]{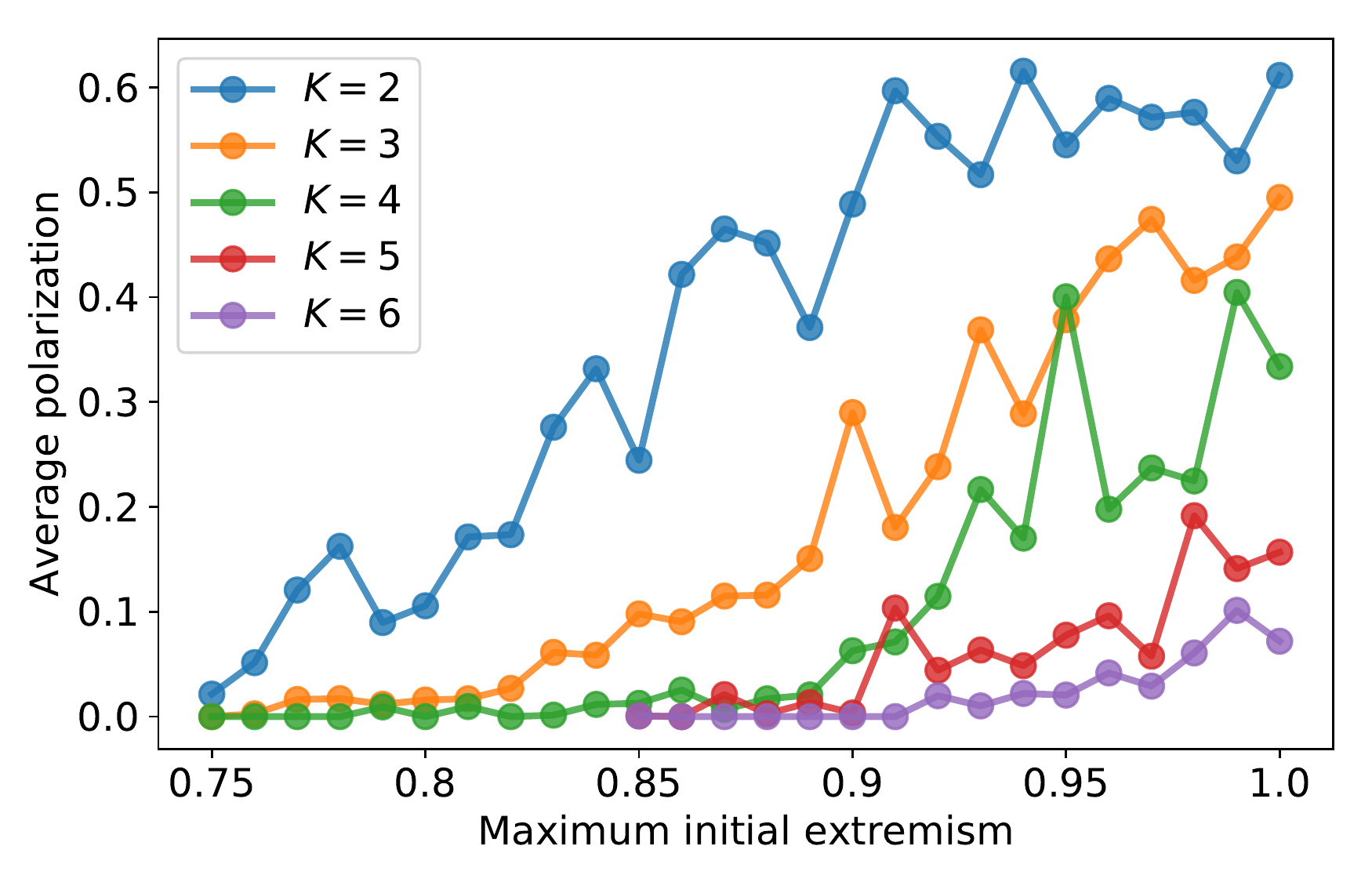}
  \caption{Average final polarization for different cultural complexities over 
    maximum initial opinion magnitude, $S$. 
    Averages are roughly zero for $S<0.75$ for all cultural complexities.
  }
  \label{fig:S_average}
\end{figure}

\begin{figure}[H]
  \centering
    \includegraphics[width=0.65\textwidth]{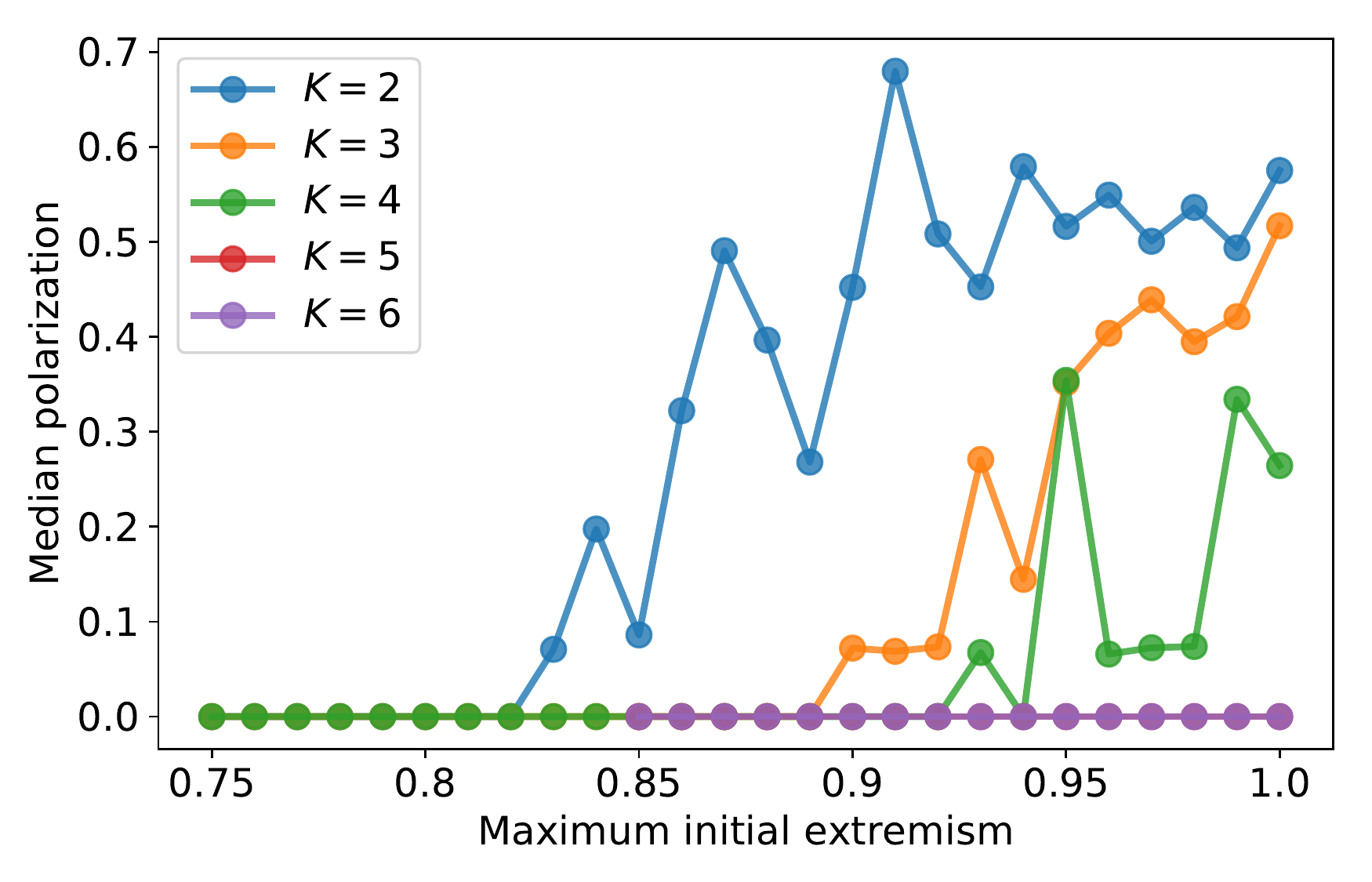}
  \caption{Median final polarization for different cultural complexities over
    maximum initial opinion magnitude, $S$.
    Median polarization for $K=5$ and $K=6$ are both flat at zero; $K=5$ 
    data is obscured by $K=6$.  
  }
  \label{fig:S_median}
\end{figure}

\begin{figure}[H]
  \centering
  \begin{subfigure}[t]{\textwidth}
    \centering
    \includegraphics[width=0.5\textwidth]{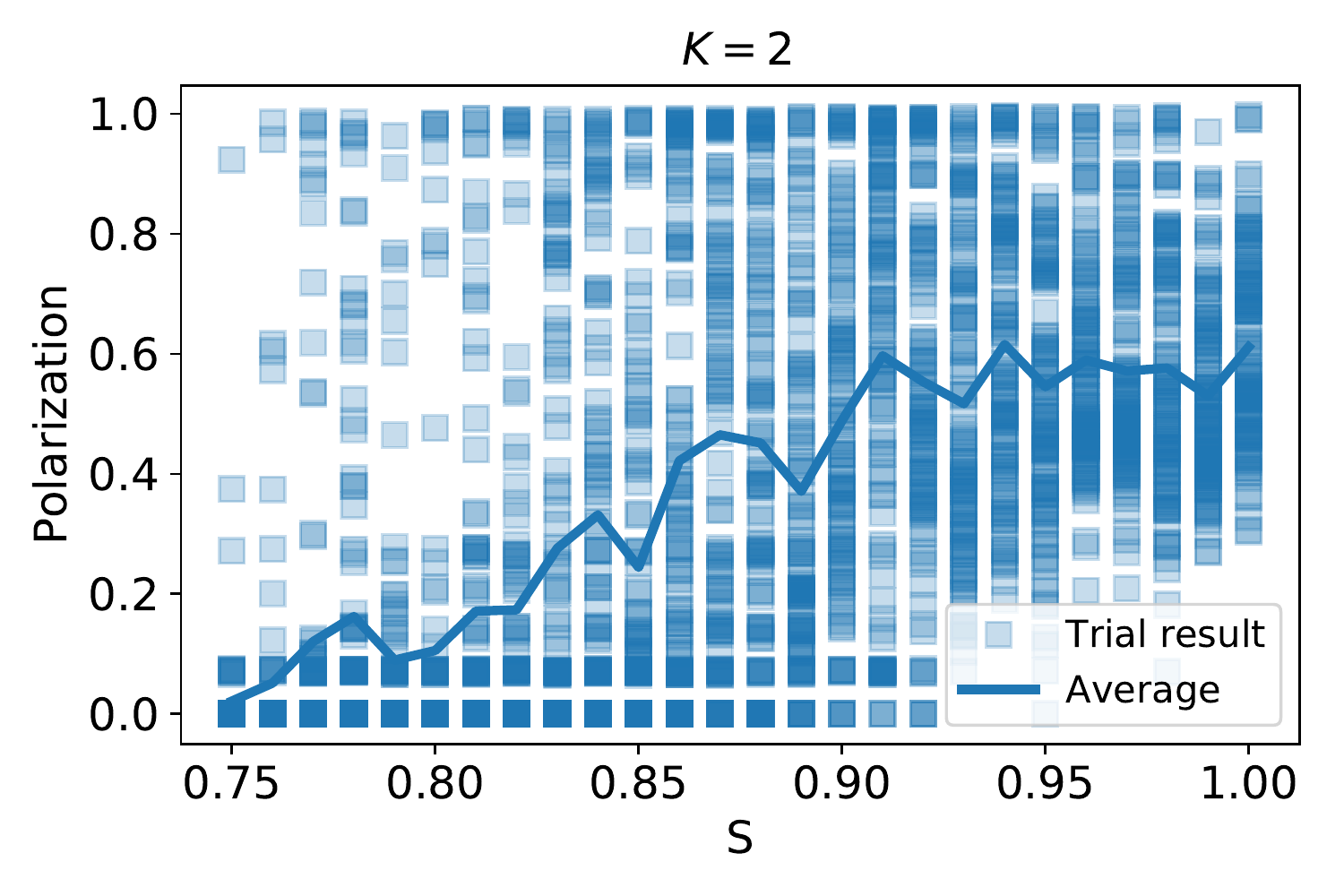}
  \end{subfigure} \\
  \begin{subfigure}[t]{0.49\textwidth}
      \centering
      \includegraphics[width=\textwidth]{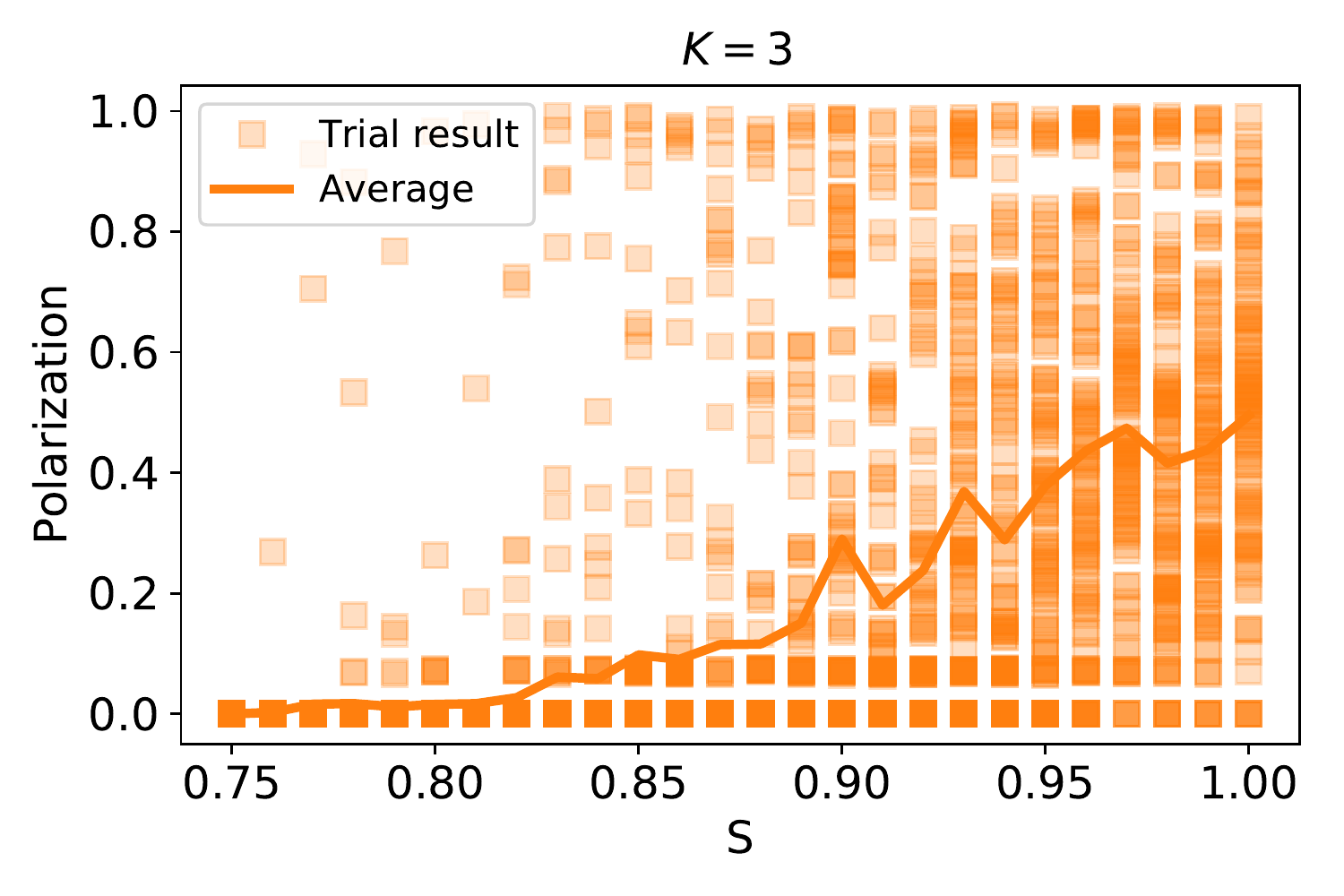}
  \end{subfigure}
  ~
  \begin{subfigure}[t]{0.49\textwidth}
      \centering
      \includegraphics[width=\textwidth]{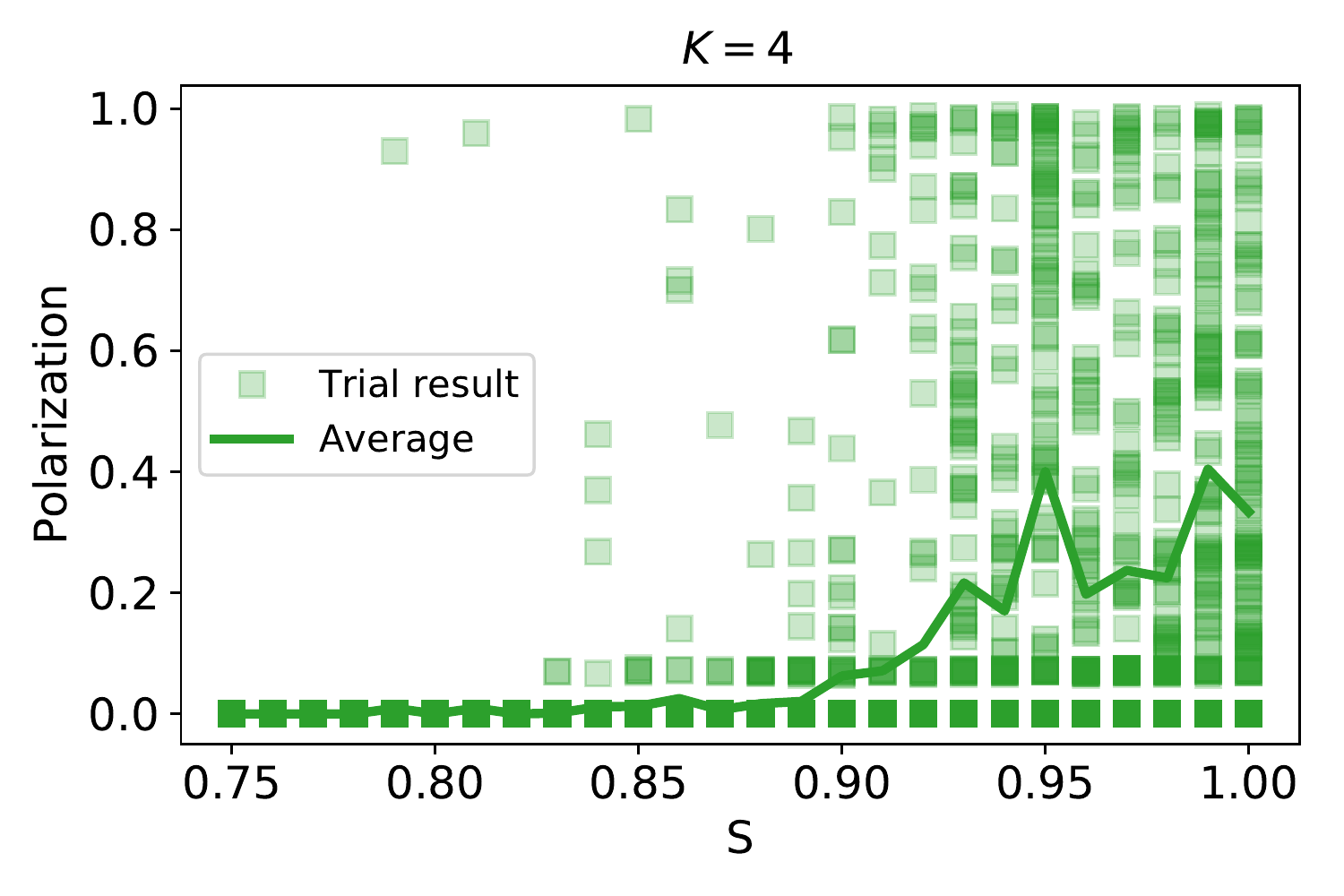}
  \end{subfigure} \\
  \begin{subfigure}[t]{0.49\textwidth}
      \centering
      \includegraphics[width=\textwidth]{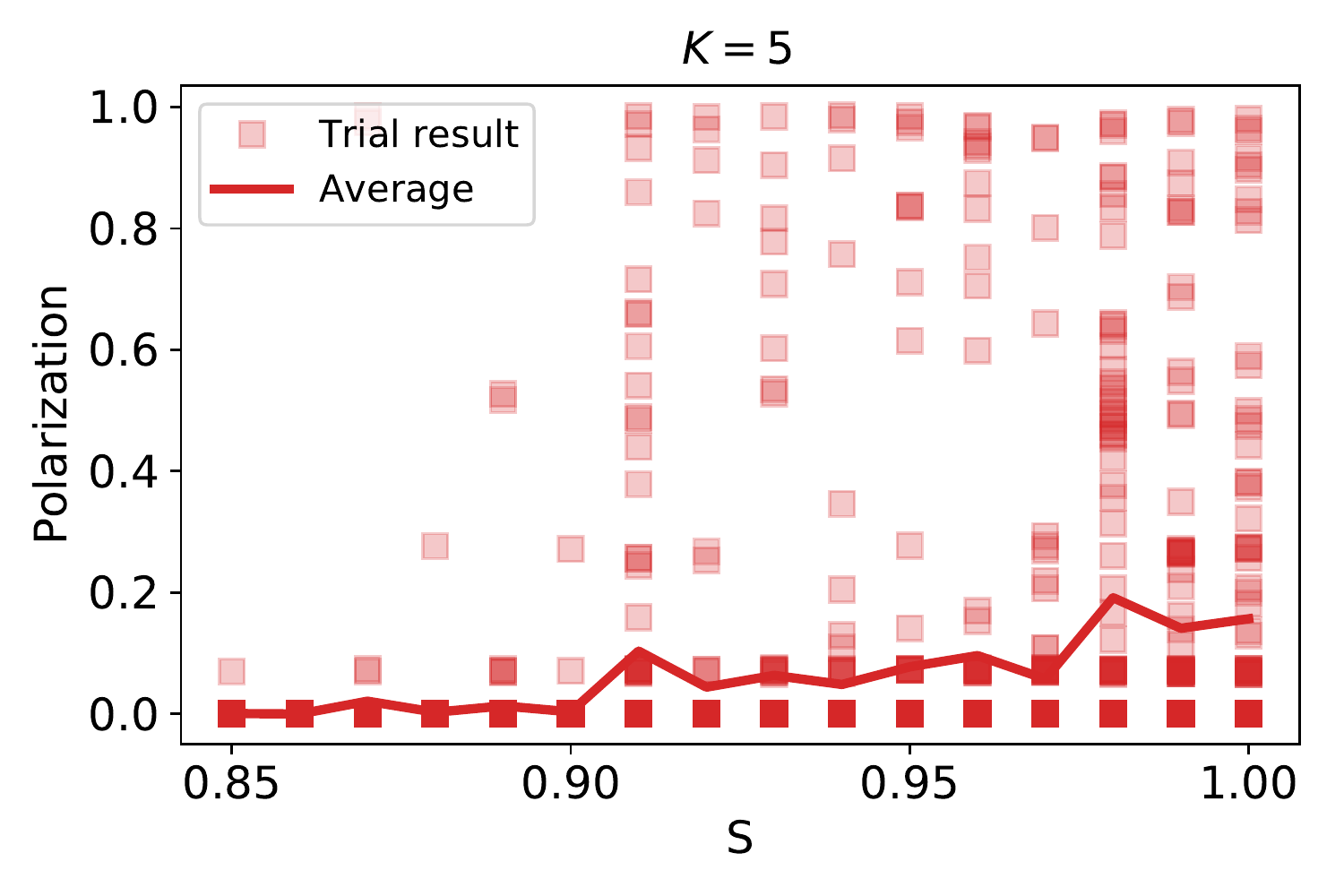}
  \end{subfigure}
  ~
  \begin{subfigure}[t]{0.49\textwidth}
      \centering
      \includegraphics[width=\textwidth]{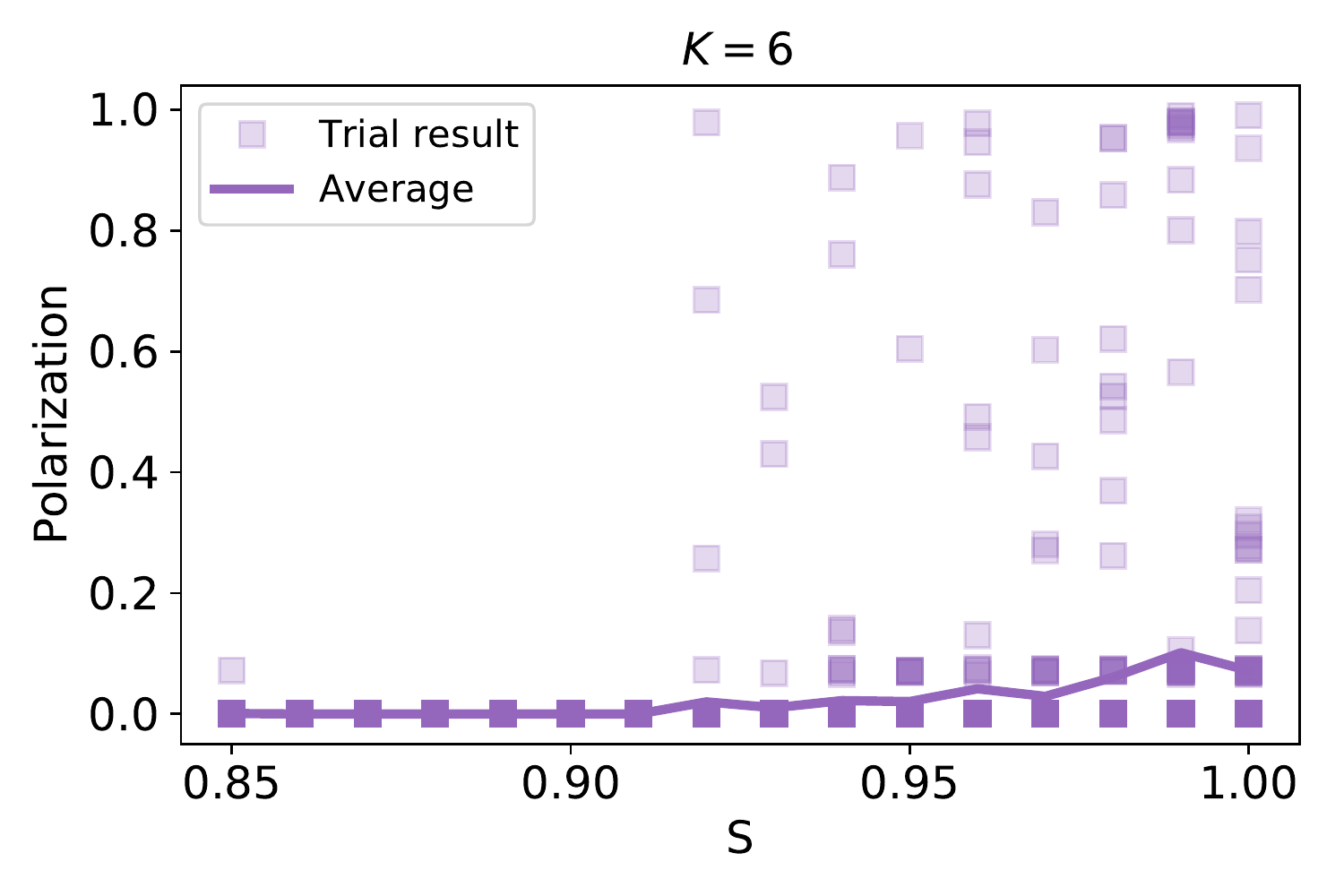}
  \end{subfigure}
  \caption{Final polarization of individual trial runs and averages from
    Figure~\ref{fig:S_average}. 
  }
  \label{fig:single_S_K}
\end{figure}

Next we extend our analysis of initial conditions further, by studying the breadth of opinions initially present in the population. 
Specifically, initial opinions were drawn from the uniform distribution $U(-S, S)$. 
Figures~\ref{fig:S_average} and \ref{fig:S_median} show the mean and median polarization of the population as function of $S$, for $K=2,\ldots,6$. 
In general, the average final polarization decreased with smaller $S$ for all values of $K$. The lines are not perfectly smooth due to the large variation in outcomes described in the previous section (see Figure~\ref{fig:single_S_K}).

We again examined the within-condition variation in final polarization (Figure~\ref{fig:single_S_K}). Even when the average polarization was very small, we nevertheless saw instances of strongly polarized outcomes for $S < 1$ across all values of $K$. For small values of $S$, much more polarization occurred with small $K$. 
This further highlights the fact that initial conditions, in conjunction with the cultural complexity, bias the system towards larger or smaller levels of polarization, but do not eliminate the possibility of either conformity or extreme polarization.

\subsection{The meaning of polarization in high-dimensional opinion space}
Clearly extreme positions are important in the FM model. Extremists are
more stubborn (and therefore more influential) than centrists due to smoothing. Our analysis indicates that under a wide range of conditions, all opinions are likely to end up at extreme values. Indeed, the only stable states of the model are complete consensus, which can be at any point in opinion space in the absence of noise, or for all opinions to be at extreme values. This brings us back to a key result of the FM model, which is that increased cultural complexity, $K$, decreases polarization. Recall that polarization is measured as the variance among distances between agent opinions. To what extent is this decrease in polarization with increased cultural complexity driven by the fact that, for larger $K$, there are simply more ``corners'' (extreme opinion values) for agent opinions to settle on? 

We investigated this question 
by comparing polarization emerging from the dynamics of the FM model with
polarization that occurs when agents are artificially placed on a random vertex of 
the $K$-dimensional opinion hypercube. We found the polarization for this
combinatorial condition is $P_{c} \approx 1/K$ via Monte Carlo 
sampling with 100 agents and 1000 trials for each $K \in \{1, \ldots, 12\}$.  
In the Appendix we derive a formal proof that $P_{c} = 1/K$ exactly in the limit as $N \rightarrow \infty$.

\begin{figure}[H]
  \centering
    \includegraphics[width=0.75\textwidth]{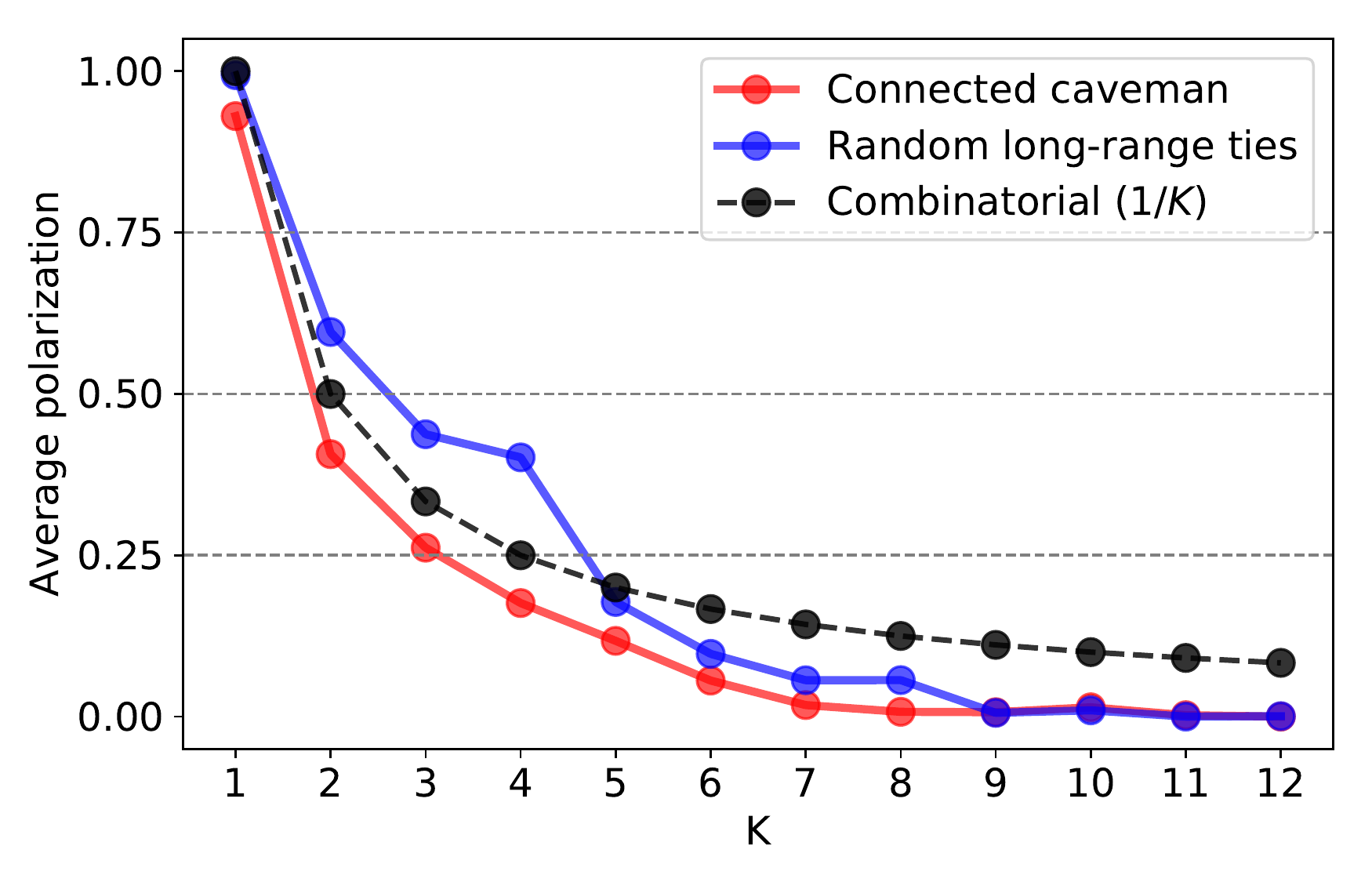}
  \caption{Polarization resulting from FM model simulations under connected caveman and random long-range tie conditions, compared with polarization resulting from agents arbitrarily choosing a corner of opinion space at random. Monte Carlo simulations revealed that polarization goes as $1/K$ if agents
    simply pick a corner at random. Random long-range and connected caveman
    data points are averaged from 100 trials with $10^4$ iterations. 
    Combinatorial condition data points are the average over 1000 trials 
    and $10^4$ iterations. Standard deviation around combinatorial trial averages was
    less than $10^{-2}$.
  }
  \label{fig:combinatorial-comparison}
\end{figure}

When we compare the combinatorial result to the FM model results, 
we find that observed decrease in polarization with increased $K$ follows the combinatorial results very closely (Figure~\ref{fig:combinatorial-comparison}).
The connected caveman condition results in a lower polarization, on average,
than $P_{c}$ for all $K$ that we tested. The random long-range condition results in an 
average polarization roughly equal to $P_{c}$ for $K=1$, higher average polarization 
than $P_{c}$ from $K=2$ to $K=4$, and lower polarization for $K \geq 5$. The
source of this jump from above-combinatorial to below-combinatorial is not
clear, but is an interesting avenue for future work.

\subsection{Noisy communication increases polarization, particularly in the
absence of initially extreme opinions}

Up to this point, we have assumed that agents accurately express their own opinions and accurately receive information concerning the opinions of others. As this assumption is unlikely to fully hold in most cases of human interaction, it is important to assess the model's robustness to noisy communication. 
To do this, we introduced random error into the opinion update equation, so that every cultural feature communication
channel, for every connected dyad, was modulated by a noise term, $\epsilon$, 
drawn from a normal 
distribution with mean 0 and standard deviation $\sigma$. 
Let us call $\sigma$ the ``noise level.'' 
We varied the noise level from 0 to 0.2 in increments of 0.02. For each of these
noise levels, we also varied $S$ from 0.5 to 1.0 in steps of 0.05 for a total of
121 parameter pairs for each  $K \in \{2, 3, 4, 5\}$. 
Note that we did not explicitly bound opinion components in the presence of noise. This led to us discarding 19 of the 60500 runs due to runaway opinions that diverged to infinity, and this was only for the highest noise levels used. These  (discarded runs had noise levels of .18 or .2). Most parameter settings had only one discarded run if any, with one parameter setting having three discarded runs, lowering the number of samples to 97 from 100 for that parameter setting (K = 5, S = 0.95, and noise level= 0.2).  This lack of smoothing had no effect on non-divergent model runs polarization outcomes, as polarization was less than or equal to 1.0 for all.

\begin{figure}[H]
  \centering
  \includegraphics[width=\textwidth]{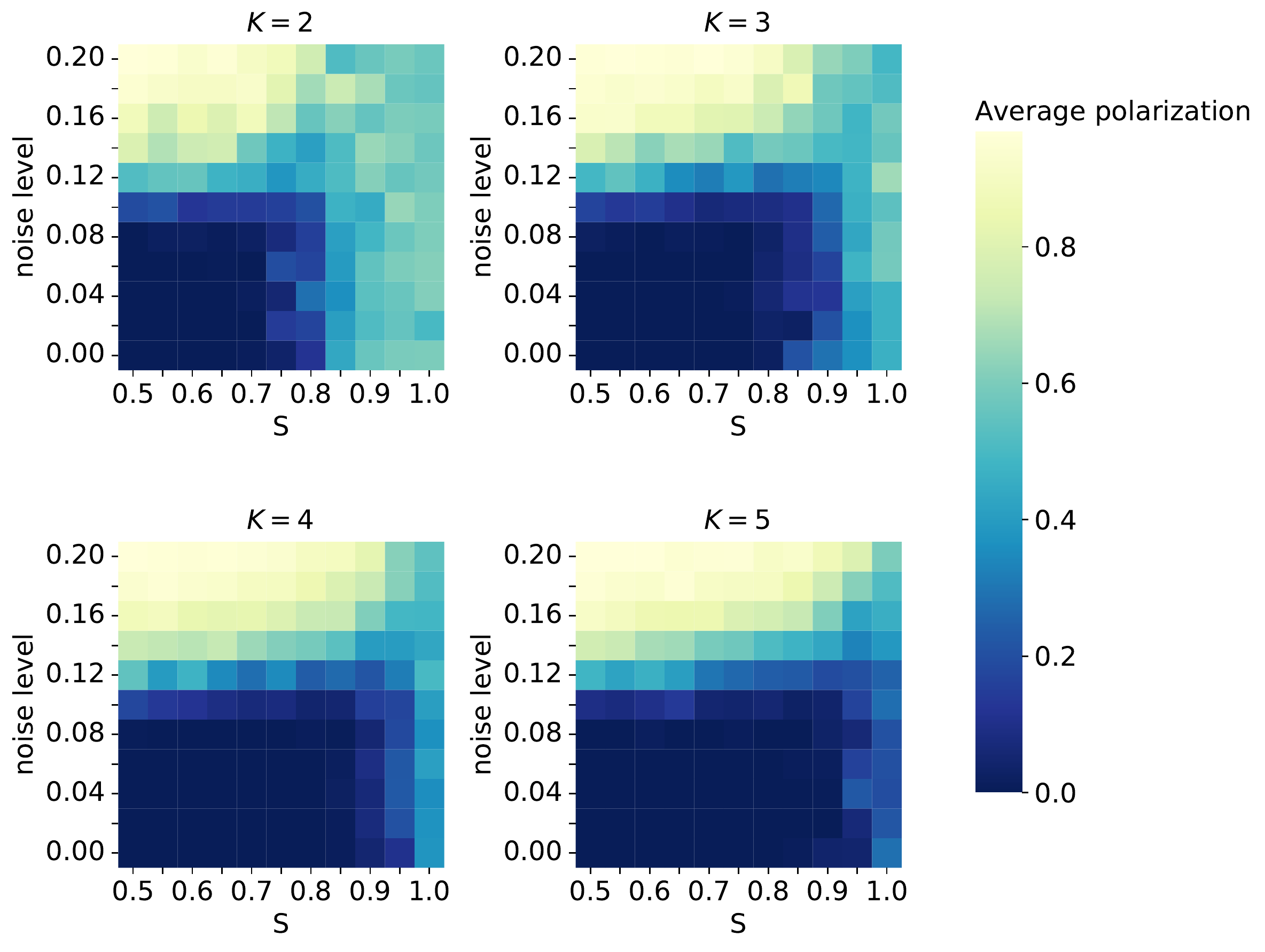}
  \caption{Final average polarization varies with both the width of the
    uniform distribution of initial opinion magnitudes and the noise level in
    the opinion updates. The value in each square of the heatmap is the average of
    100 trials. 
  }
  \label{fig:heatmaps}
\end{figure}

\begin{figure}[H]
  \centering
      \begin{subfigure}[t]{0.49\textwidth}
          \centering
          \includegraphics[width=\textwidth]{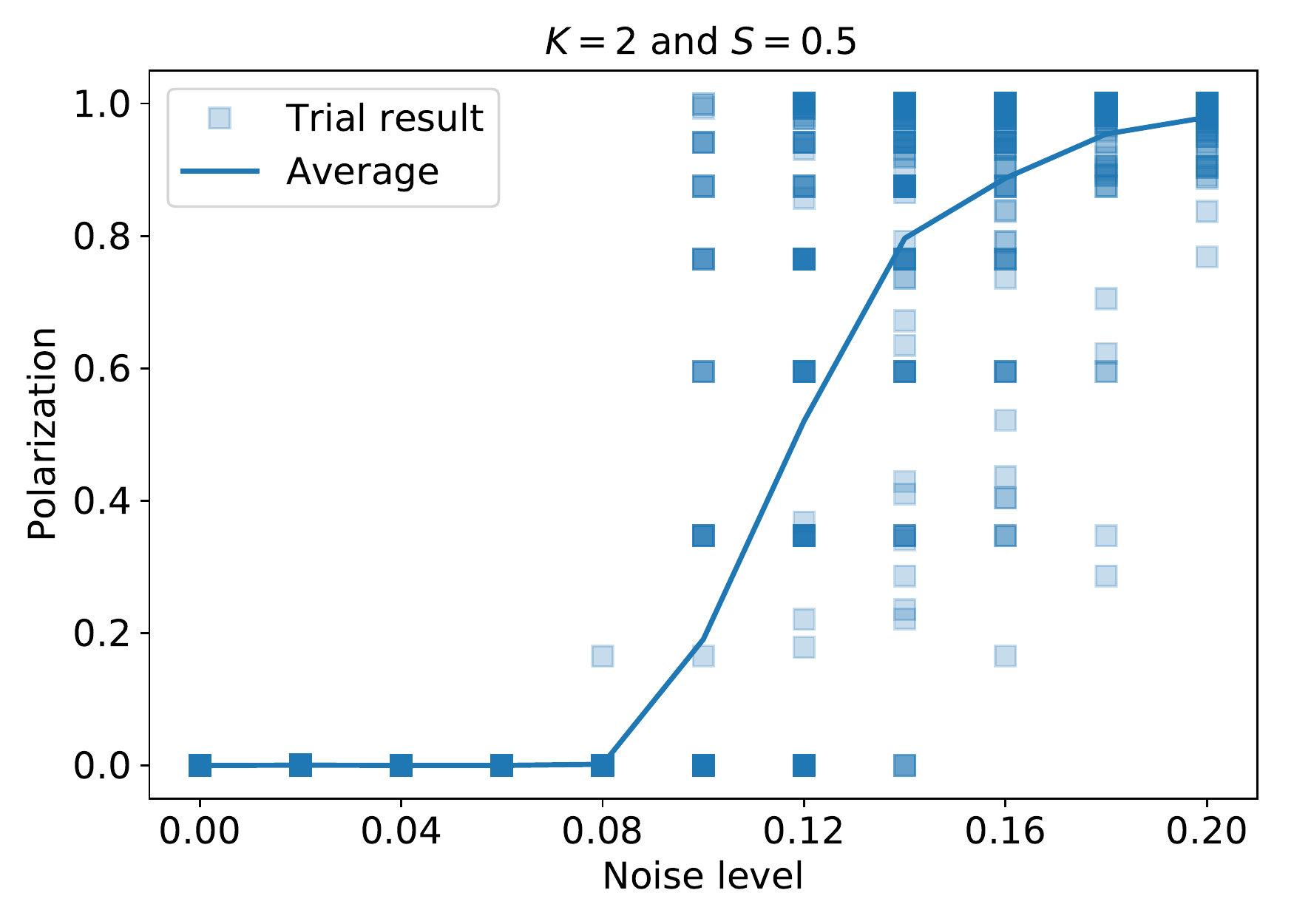}
      \end{subfigure}
      ~
      \begin{subfigure}[t]{0.49\textwidth}
          \centering
          \includegraphics[width=\textwidth]{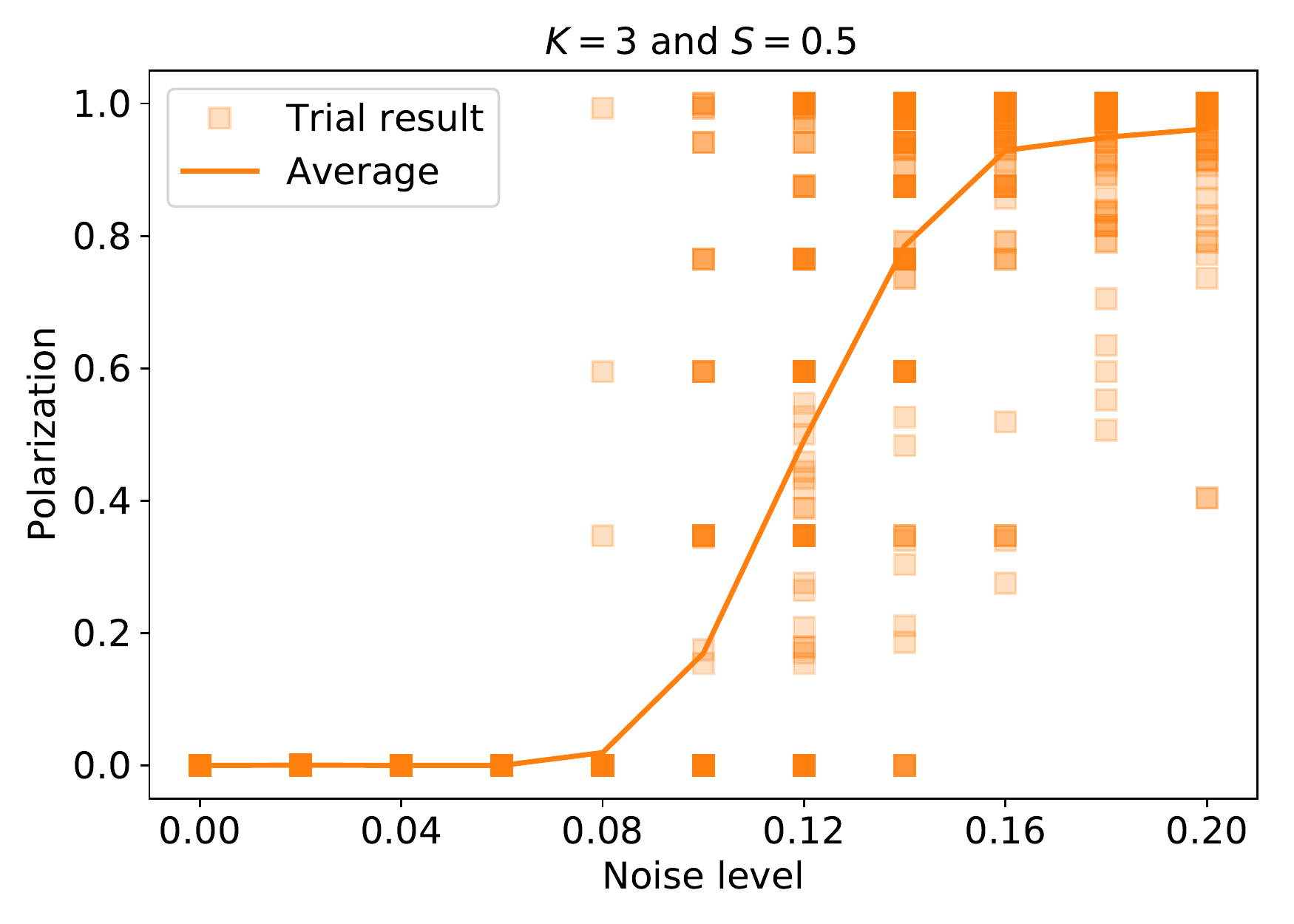}
      \end{subfigure} \\
      \begin{subfigure}[t]{0.49\textwidth}
          \centering
          \includegraphics[width=\textwidth]{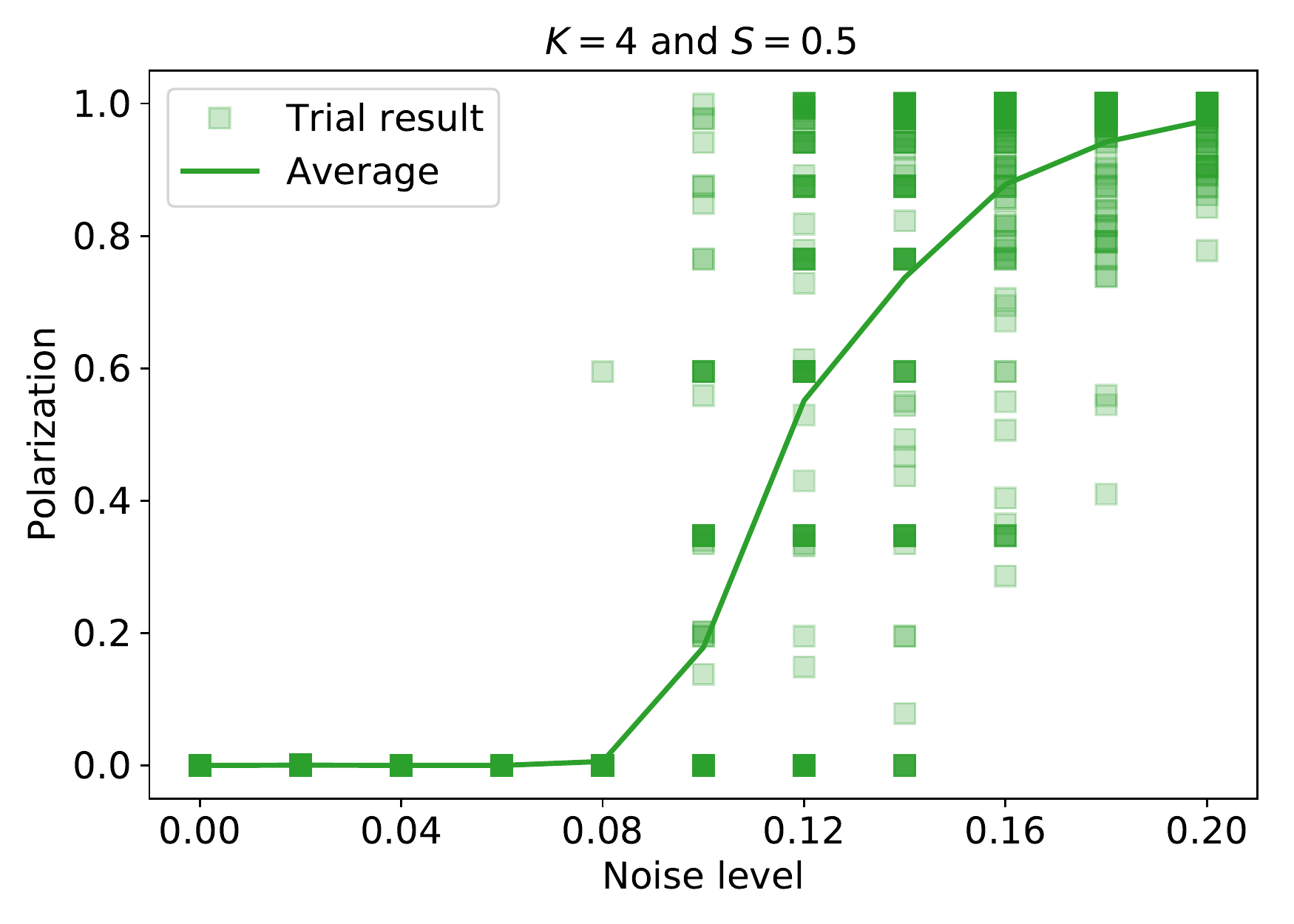}
      \end{subfigure}
      ~
      \begin{subfigure}[t]{0.49\textwidth}
          \centering
          \includegraphics[width=\textwidth]{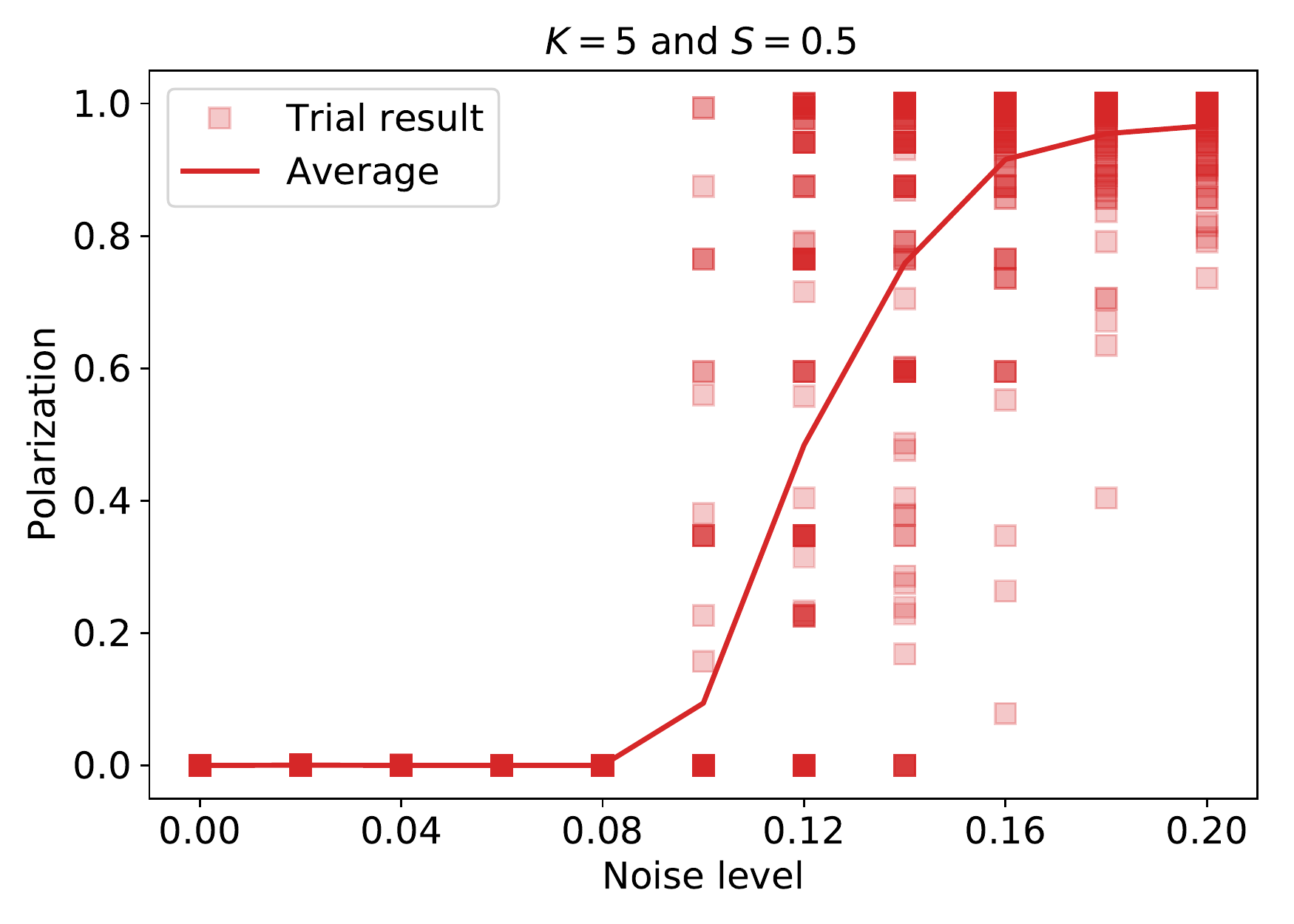}
      \end{subfigure}
  \caption{Final polarization of individual trial runs and averages from
    Figure~\ref{fig:heatmaps} for $S=0.5$ as a function of noise level, $\sigma$. 
    As the noise level is increased, the system is increasingly biased towards larger final polarization outcomes.
  }
  \label{fig:single-runs-commnoise}
\end{figure}

These experiments reveal an interesting pattern of results.
A sufficiently large amount of noise produced high levels of polarization for low values of $S$, which never produced polarization in the absence of noise. 
Indeed, there appears to be a phase transition point for $\sigma$ under low $S$, 
below which the system collapses to complete conformity and above which 
we see high levels of polarization (Figure~\ref{fig:heatmaps}). Across the values of $K$ we tested, this 
threshold appealed to be around $\sigma=0.8$, below which we never saw any 
polarization for low $S$ (Figure~\ref{fig:single-runs-commnoise}). 
As $S$ increases, however, the system behavior becomes less sensitive to 
noise, appearing to be completely insensitive to noise close to $S=1$.

\begin{figure}[H]
  \centering
  \includegraphics[width=\textwidth]{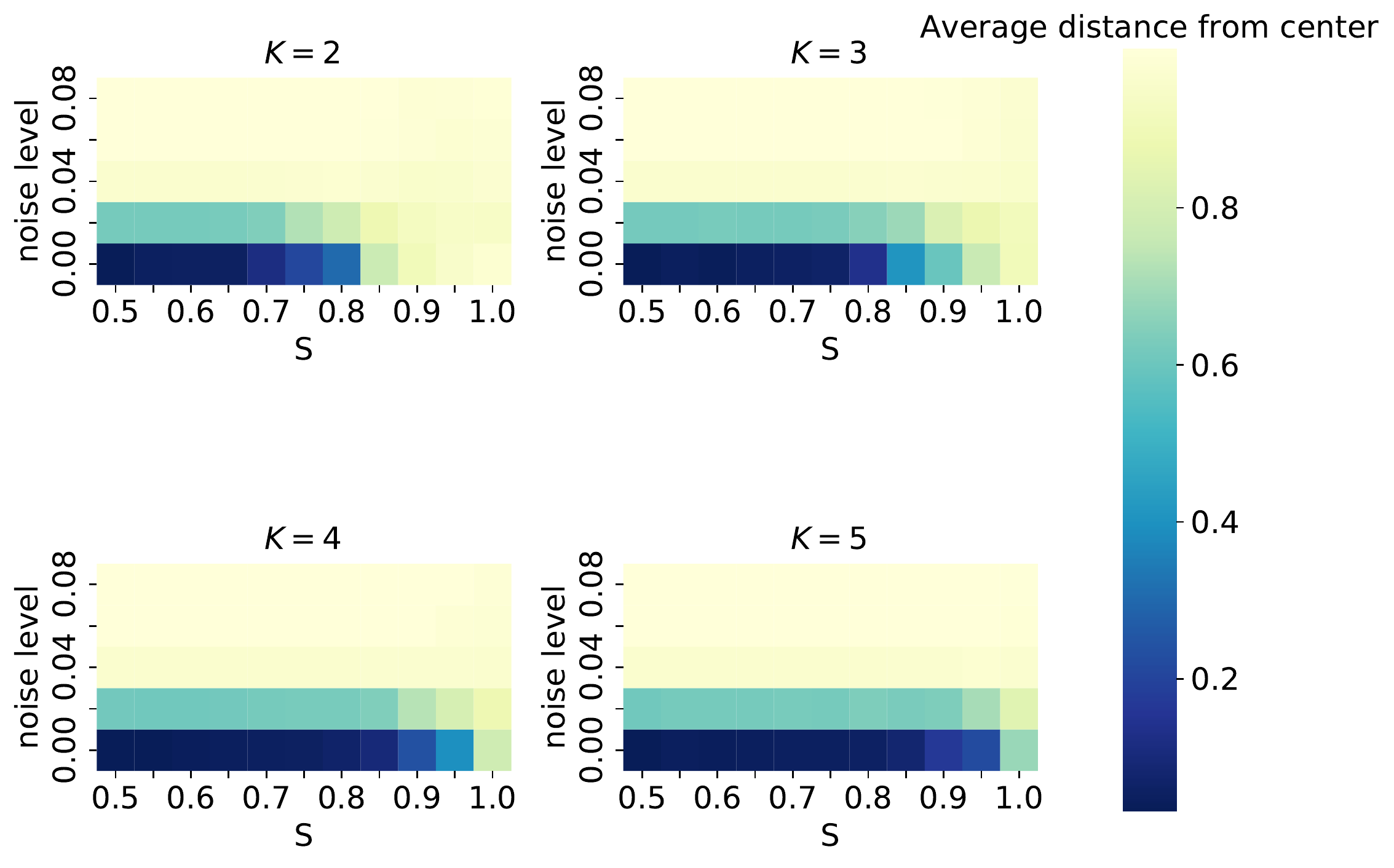}
  \caption{Noisy communication causes extremism without polarization before
    it causes extremism with polarization.  For all $K$ pictured, the
    average distance from center increases with moderate levels of noise, even
    though polarization has not increased, as shown in Figure~\ref{fig:heatmaps}.
    The value in each square of the heatmap is the average of
    100 trials. 
  }
  \label{fig:avedist_heatmaps}
\end{figure}

Even though polarization is rare at moderate noise levels, extremism is not.
A noise level of over 0.1 was required to reliably drive
the system to polarization in our simulations, but lower noise levels led to consensus around an
extreme location in opinion space rather than at a most centrist position. We infer this because the average agent distance from center increases to the maximum, 1.0, with noise levels of only 0.6
(Figure~\ref{fig:avedist_heatmaps}). Thus, we obtain the interesting result that
even small amounts of communication noise can move the population to extremist
positions.

\begin{figure}[H]
  \centering
    \begin{subfigure}[t]{\textwidth}
      \centering
      \includegraphics[width=.7\textwidth]{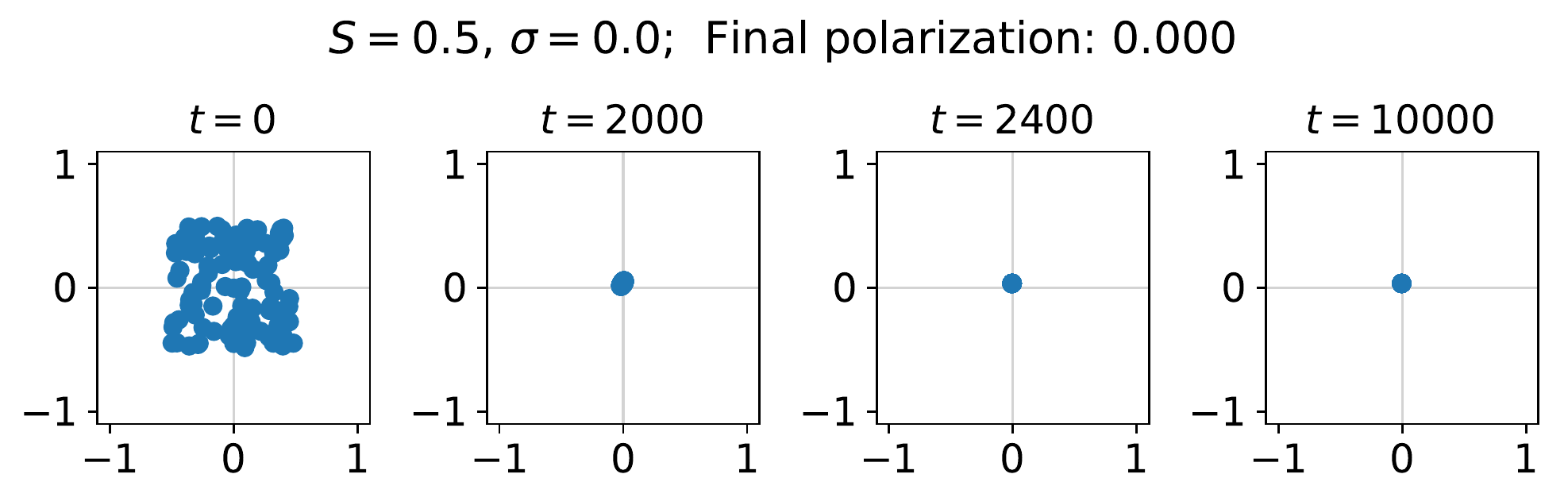}
    \end{subfigure} \\
    \begin{subfigure}[t]{\textwidth}
      \centering
      \includegraphics[width=.7\textwidth]{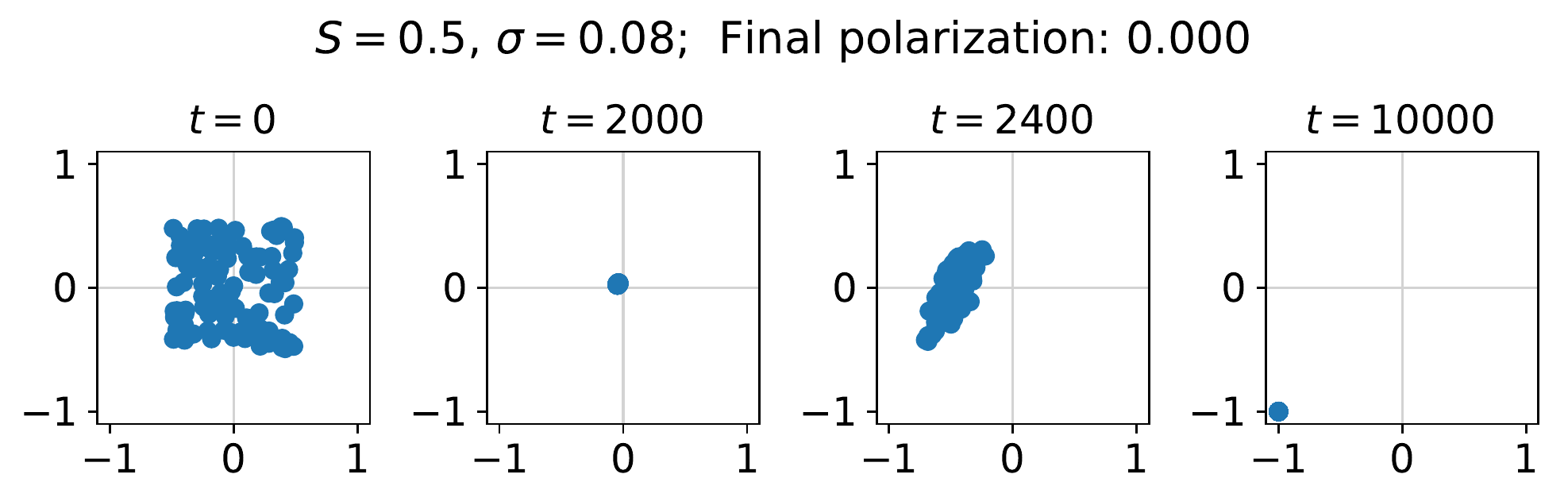}
    \end{subfigure} \\
    \begin{subfigure}[t]{\textwidth}
      \centering
      \includegraphics[width=.7\textwidth]{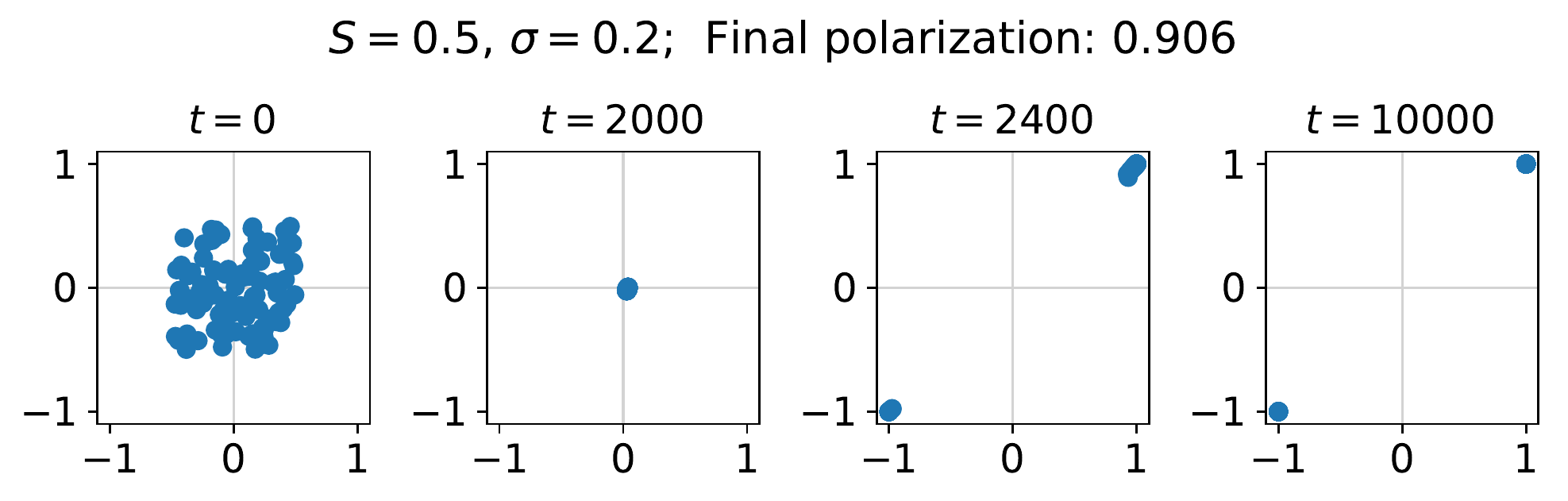}
    \end{subfigure} \\
  \caption{Exemplar spatiotemporal dynamics of agent opinion coordinates with $K=2$ and 
    $S=0.5$ for $\sigma \in \{0.0,0.08,0.2\}$. There are three regimes. In the first, without
    noise, every simulation 
    ends in centrist consensus (top row). In the presence of noise with $\sigma=0.08$,
    agents find extremist consensus; in this trial agents found consensus around the point $(-1, -1)$.
    The third regime is the high polarization regime at the highest level of communication noise
    we tested, $\sigma=0.2$. In this regime, agents split into 
    opposing camps, led by first-mover extremists.
  }
  \label{fig:noise_coords_S0p5}
\end{figure}

Figures \ref{fig:heatmaps} and \ref{fig:avedist_heatmaps} also illustrate a curious interaction between noise level, $\sigma$, and initial extremism, $S$. For smaller $S$, we observe clear phase transitions from centrist conformity to extremist conformity to polarization. For larger $S$, the populations responses are less clearly delineated.
To help explain, we present illustrations of the spatiotemporal dynamics of the model for exemplar trials. 
Consider first a case of very low initial extremism, $S=0.5$  (Figure~\ref{fig:noise_coords_S0p5}). 
In the absence of noise, the system collapses around the center of opinion space at $t=200$, and by 
$t=3000$ has reached full consensus (Figure~\ref{fig:noise_coords_S0p5}, top row). At the other extreme, under high levels of noise, $\sigma=0.2$,
agents reach a near-consensus by $t=1000$ and remain there until $t=2000$, when random long-range ties are added. 
At this point, agents are exposed to individuals with very slightly different sets of opinions, and those differences are amplified by the noise, leading to repulsion. This is sufficient to jolt the system away from conformity and into opposing camps moving towards opposing corners 
(Figure~\ref{fig:noise_coords_S0p5}, bottom row). 

For $\sigma=0.08$ we found most simulations end in extreme consensus. That is, all opinions were at the extremes ($\pm 1$) rather than closer to zero, but these opinions were universally shared so that final polarization was zero. One such trial is shown in the middle row of Figure~\ref{fig:noise_coords_S0p5}.This occurs because noise is sufficient to move the population toward the extremes (from which it is  difficult to return to center), but agents remain sufficiently clustered so that all forces remain attractive rather than repulsive.

\begin{figure}[H]
  \centering
    \begin{subfigure}[t]{\textwidth}
      \centering
      \includegraphics[width=.7\textwidth]{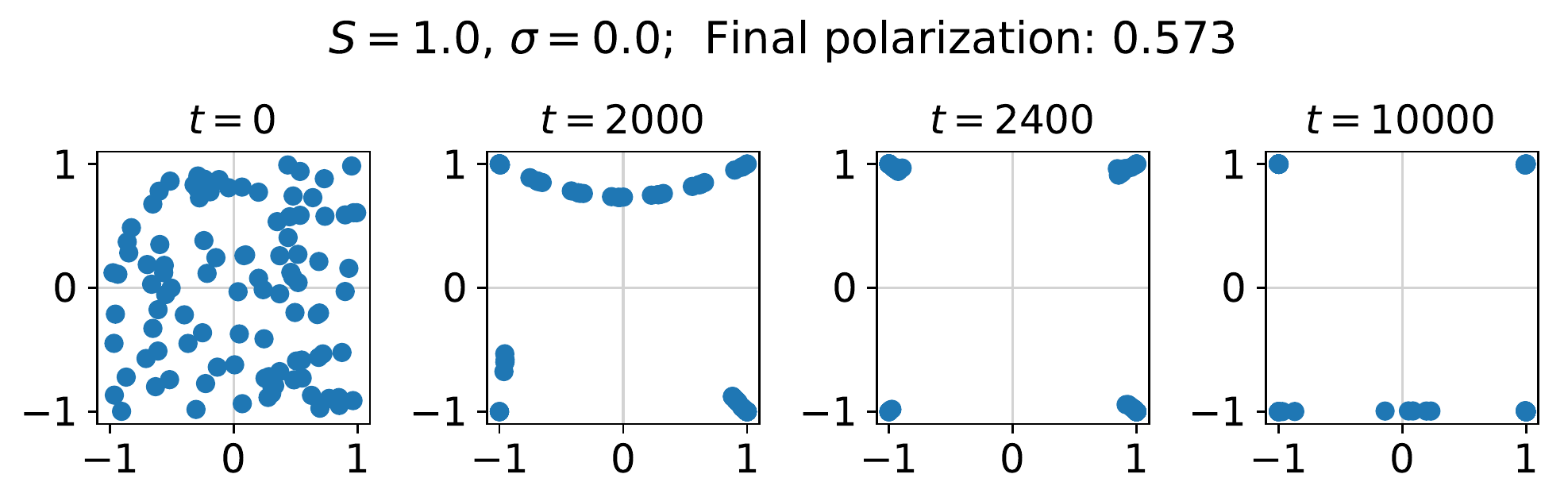}
    \end{subfigure} \\
    \begin{subfigure}[t]{\textwidth}
      \centering
      \includegraphics[width=.7\textwidth]{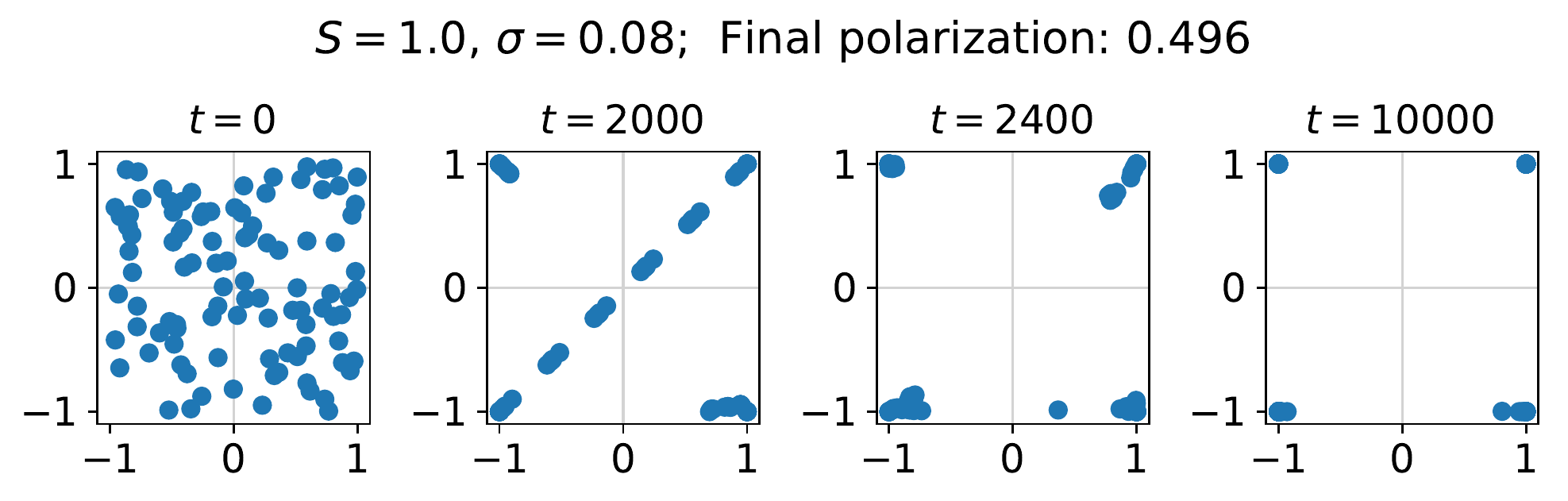}
    \end{subfigure} \\
    \begin{subfigure}[t]{\textwidth}
      \centering
      \includegraphics[width=.7\textwidth]{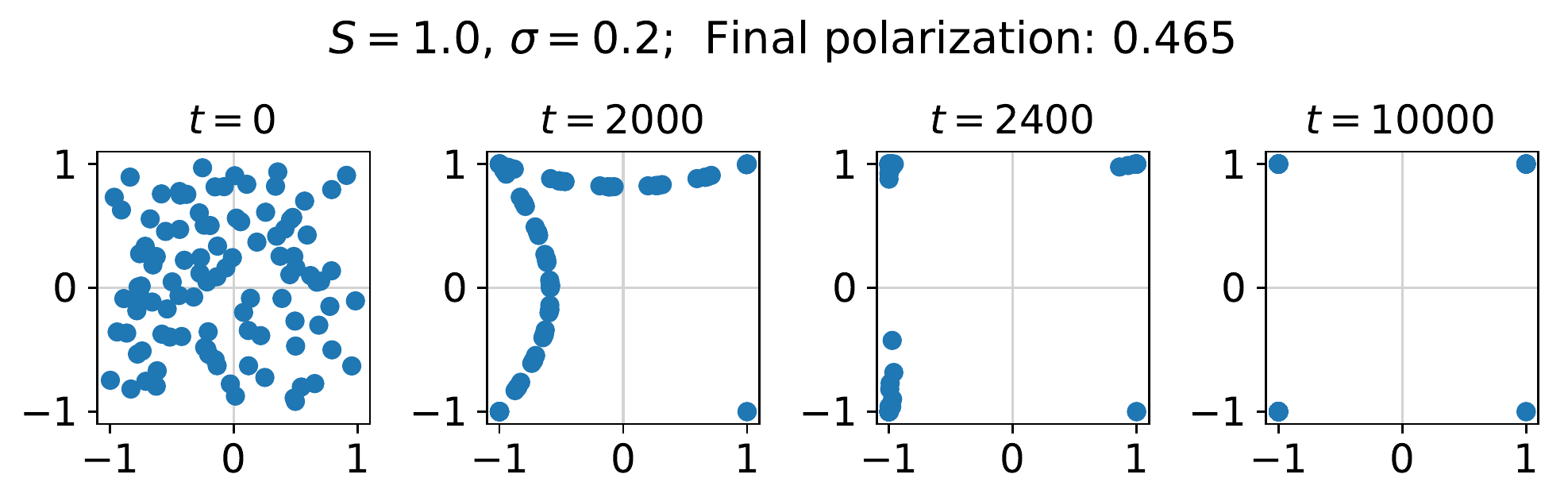}
    \end{subfigure} \\
  \caption{Exemplar spatiotemporal dynamics of agent opinion coordinates with $K=2$ and
    $S=1.0$ for $\sigma \in \{0.0,0.08,0.2\}$.  Before the random long-range ties are added at 
    $t=2000$, extremists pull centrists to the extremes, but more centrist
    agent caves are balanced between more extreme caves. When long-range ties
    are added, the balance is broken and agents proceed to move to one of the
    extremes. Because at least some extremists held each of the corners, 
    centrist agents do not move only to polar opposite corners, but in many 
    cases to the nearest corner contained a neighboring (in the network sense) agent. 
  }
  \label{fig:noise_coords_S1p0}
\end{figure}

\begin{figure}[H]
  \centering
    \begin{subfigure}[t]{0.48\textwidth}
      \centering
      \includegraphics[width=.95\textwidth]{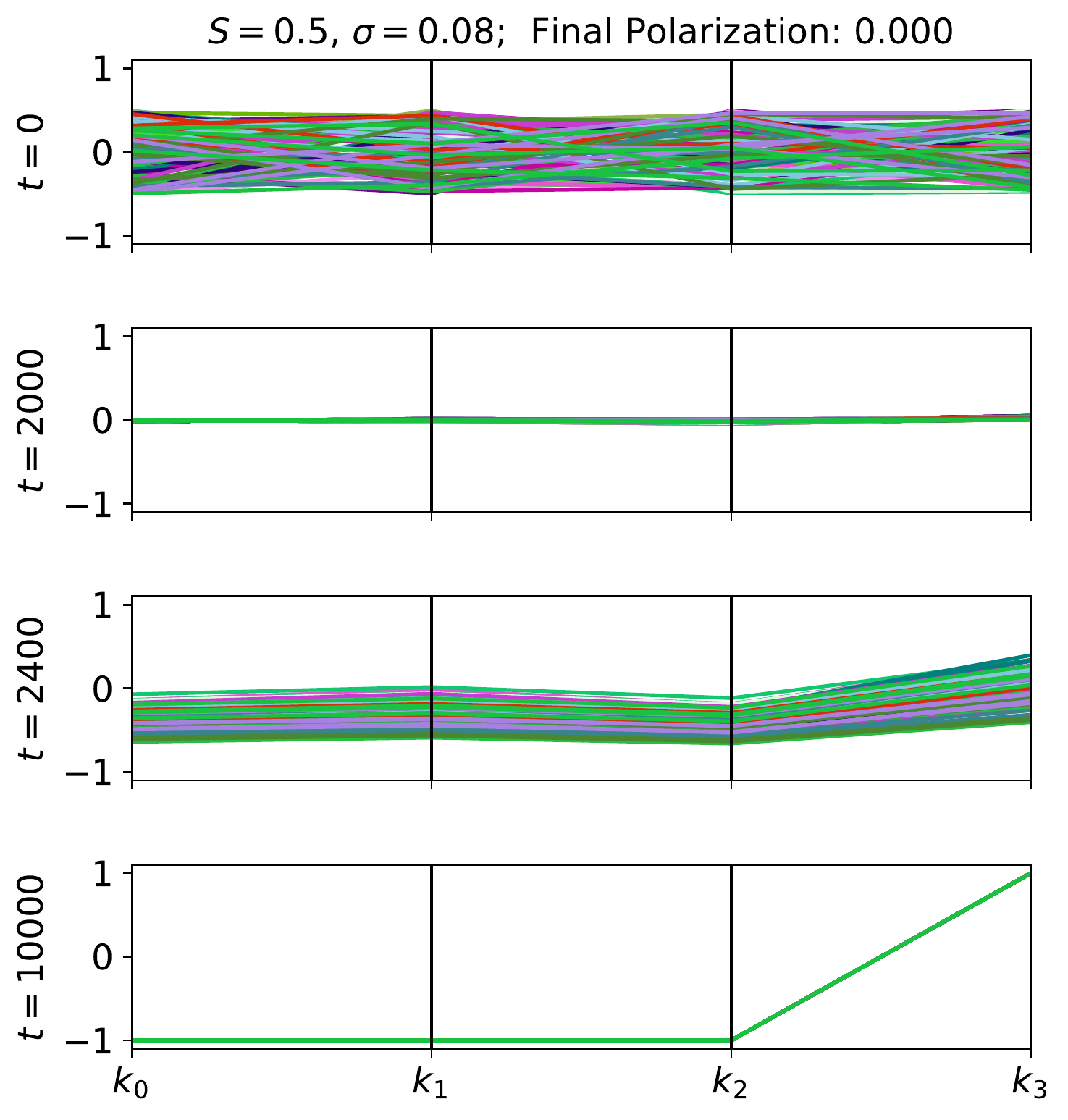}
      \caption{Moderate noise, extreme consensus}
      \label{fig:noise_coords_S0p5_K4_noise0p1}
    \end{subfigure}
    \begin{subfigure}[t]{0.48\textwidth}
      \centering
      \includegraphics[width=.95\textwidth]{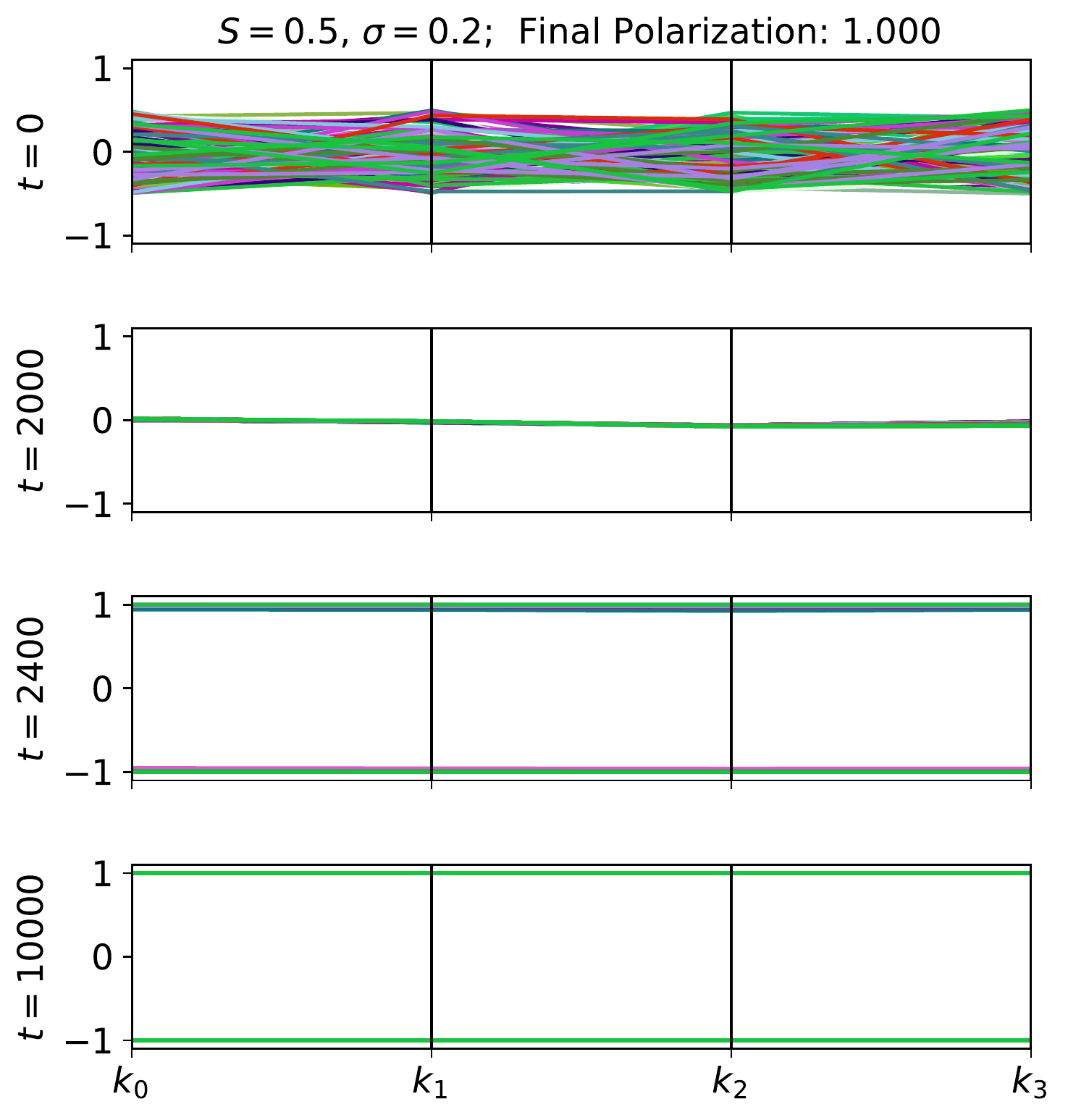}
      \caption{More noise, polarization}
      \label{fig:noise_coords_S0p5_K4_noise0p2}
    \end{subfigure} 
  \caption{Exemplar parallel coordinate timeseries for $K=4$ and $S=0.5$. 
    Here the x-axis represents a single opinion coordinate, $k_i$, and the 
    y-axis is the location of an agent for that coordinate. Each agent is
    represented by a line, colored by cave membership.
    With $\sigma=0.1$, 
    consensus emerges but at a corner of the opinion space. 
  }
  \label{fig:noise_coords_S0p5_K4}
\end{figure}

When initial opinions are drawn from the full range of possibilities ($S=1$), the system always achieves some degree of polarization. Because noise only serves to increase the likelihood of extreme opinions, this condition is unaffected by noise. 
 Typical cases are shown in Figure~\ref{fig:noise_coords_S1p0}. The behavior for
$t\leq 2000$ is similar in all three cases: each cave reaches a local consensus,
and the network of caves reaches a stable configuration. 
Some of the caves find consensus values at the corners. When
random ties are added, the stable configuration is broken, and agents are
pulled towards one of the four corners, where some caves have already 
been stably established. The caves in the corners do not move. Recall that a key assumption of the FM model is that extremist opinions influence centrist opinions more than centrists influence extremists. 
The noise is not strong enough to move extremists from extreme positions. In other words, in the presence of extreme opinions, network structure, not noise, dominates the dynamics.
We extend the intuition to higher dimensions of opinions space using parallel coordinate plots, visualizing time series of opinion dynamics for $K=4$ (Figure~\ref{fig:noise_coords_S0p5_K4}). 

\section{Discussion}

Humans are the quintessential cultural species. Our instinct to learn from others is a key reason for our domination of the planet \cite{henrich2015secret,laland2017darwin}. An under-appreciated component of cultural learning concerns exacerbating differences and rejecting opinions when individuals are not likely to share one's current norms and beliefs. When those differences occur within a community, they can lead to discord. Many of us live in multicultural societies requiring cooperation and common ground, and so it natural to ask: when do we expect polarization, and is there anything we can do about it. Any suggestions based on our modeling efforts here should of course be compared with empirical studies. Hopefully these results stimulate further 
empirical work to understand when and why polarization emerges in
real-world situations. One such opportunity for future work is to connect our findings to 
the political science literature on polarization \cite{Sides2015}, 
especially in relation to communication. 
If agents had different roles, such as elite agents (politicians and media) and common agents, 
we could model the effects of ideologically-biased news in political polarization \cite{Prior2013,Pew2014}. 
Our results show that in the presence of sufficiently large communication noise and 
small-world networks, a situation we are arguably in today, a state of polarization 
is the only stable state (Figures~\ref{fig:noise_coords_S0p5},~\ref{fig:noise_coords_S1p0}, and~\ref{fig:noise_coords_S0p5_K4}). 
It is interesting to consider this in light of one recent analysis suggesting that the United States Constitution was designed not just to accommodate polarization, but to foster it for the sake of stability 
\cite{Wood2017b}. 

We have highlighted the stochastic nature of the system being modeled. A key conclusion is that empirical results of opinions on social networks may, when taken on a case-by-case basis, exhibit trends that bear little resemblance to those predicted by the model. This is not necessarily an invalidation of the model, but merely a consequence of the variability inherent in complex systems. That said, given enough data, key trends should emerge.  
We have confirmed Flache and Macy's (2011) result that long-range ties increase polarization. As such, we might emphasize the importance of local communities being allowed to reach their own consensus. We have shown that decreasing initial extremism can reduce polarization, as one might expect. Achieving consensus in a community relies heavily on the absence of opinions at the extremes. 
However, this result is quite sensitive to noise in communication. A little bit of noise can shift consensus from centrist or ambivalent positions to more extreme views, while more noise can lead to polarization. 
Even if polarization is to be avoided, what about the intermediate case of 
``extreme consensus''? While it may be natural to view extreme opinions as 
undesirable, an alternative perspective is that they represent a more stable system of 
cultural coherence. Note that these findings contradict computational and mathematical studies of the 
bounded confidence model under the influence of noise, where sufficient noise breaks 
polarization and leads to disordered opinion spreading \cite{Pineda2009,Carro2013,Kurahashi-Nakamura2016}. 
This is because in the bounded confidence model, agents that are too far from 
one another do not interact. In the FM model, connected agents always interact, 
and the further apart they are in opinion space, the more strongly they repel 
one another in opinion space. 

We confirmed Flache and Macy's (2011) result that increased ``cultural complexity''---the number of  opinions that are important to individuals in assessing their similarities and differences with
others---decreased overall polarization. We also showed that this result stems directly from an increase in the number of permutations of extreme opinions individuals can hold when there are more items on which one can hold opinions. This might be viewed as a flaw in the metric of polarization used here. Alternatively, we believe it is reasonable to posit that a community with a wider diversity of views should be considered less polarized than a community with only a few suites of clustered opinions. In any case, this finding highlights the importance of a thorough understanding of one’s distance measure when dealing with multidimensional opinions. 
Our analysis may in fact cast doubt on the interpretation by Flache and Macy (2011) that 
cultural complexity decreases opinion polarization, if one also rejects the interpretation 
that adding arbitrary traits on which actors are indifferent should reduce their opinion 
distance.

As noted, the model we have studied is a simplified abstraction, and does not include many details that are important to the empirical reality of opinion dynamics. In general, theoretical modeling work should start simple, and gradually add heterogeneity as the simpler versions of the system in question become fully described. Future work should explore these sources of heterogeneity. 
First, we did not distinguish between private opinions and public productions
representing those opinions \cite{Nowak1990}. Our operationalization of communication
noise could be interpreted as a modulation of private opinion, 
but communication noise could also be interpreted as misunderstanding of perfectly-reproduced, publicly voiced opinions. People often communicate public opinions that differ from their private opinions 
when incentives for the parties involved are not aligned \cite{crawford1982strategic,pinker2008logic,smaldino2018evolution}.  
Second, we ignored the structural influence of explicit identity groups. It could be argued that clustering of agent opinions implicitly defines an identity group. For example, \citeA{DellaPosta2015}
measured network auto-correlation to explain why people's preferences cluster
together. This data-driven approach was offered as an attempt to explain arbitrary opinion clustering, as indicated by the paper's title, ``Why do liberals drink lattes?''. Nevertheless, explicit identity with groups and roles influences human behavior far beyond homophilic clustering \cite{barth1969ethnic,berger2008drives,smaldino2018social}. 
Third, we ignored individual differences in how individuals influence and are influenced. Some people may be stubborn while others are easily swayed. 
Some prestigious or charismatic individuals may have outsized influence while others are ineffective at communicating their opinions. Relatedly, individuals may also vary in their confidence in their opinions, which will influence the extent of their mutability and persuasion. 
The assumption that as agents become more extreme, their opinions become more stubborn, as formalized in Equation~\ref{eq:smoothed-update}, may not always hold. Indeed, our work highlights the need for additional empirical work on how individuals alter their opinions as a function of how extreme those opinions are.
Finally, the social networks used in our model are simplistic in both dynamics and structure. Ties in many real world networks change with greater frequency than we modeled, providing new opportunities for social influence. Moreover, interactions and opinions are contextual. Individuals are embedded in multilayered social networks, in which the dynamics of opinions may be considerably more nuanced than indicated by our relatively static, single-layer network \cite{battiston2017layered,smaldino2018resilience}. 

In our study of the FM model we have found rich behaviors and theoretical lessons for understanding opinion dynamics. This work highlights the potential for complexity even in a very simple model of individual behavior, because network structure provides for path dependent effects and can be further influenced by initial conditions and noise. Our analytic approach highlights the value of systematic investigation of a model's explicit and tacit assumptions. 

\section*{Appendix: Proof that polarization scales with $1/K$}

We hypothesized that the decrease in polarization with increasing $K$ observed in simulations of the FM model were driven by an increase in the number of permutations of binary vectors of length $K$, in which each element was $-1$ or 1. We supported this hypothesis in the main text with simulations in which agents were randomly initialized at such extreme positions in opinion space. Here we derive a formal proof that polarization in the FM model scales with $1/K$ if we assume that agents are randomly assigned a vector of ``extreme'' opinions, such that $\forall {i,k}$, $s_{ik} \in \{-1, 1\}$. 
To do this, we exactly calculate the polarization of a population 
where each agent occupies one of the $2^K$ corners of opinion space with $K$
cultural features. 

Recall that polarization is defined as the variance in pairwise distances
between all agents. We define the \emph{combinatorial polarization}, $P_c(K)$, 
as the polarization that arises from randomly placing each agent at one of the $2^K$
corners of opinion space with $K$ cultural features, 
which is a $K$-hypercube, denoted $Q_K$. ``Corners'' of opinion space are 
simply vertices in the graph of $Q_K$.
We computed this value numerically for $K \in \{1, \ldots, 12\}$ and found it tracks
closely to $1/K$ (see Figure~\ref{fig:combinatorial-comparison}). 
Here we demonstrate that $P_c = 1/K$ exactly in the large $N$ limit.
To calculate $P_c(K)$, we need three elements. First, we need to
calculate the distance between pairs of agents at different corners of
$Q_K$. Second, we must count the number of agent pairs separated by the distance
from one corner to another. We do this by first counting the number of 
subcubes of dimension $L$, or $L$-subcube.
Then we count the number of maximally separated pairs in a subcube of 
$L$-subcube. Finally, we calculate the distance of maximally separated, 
or \emph{antipodal} pairs, of agents in an $L$-subcube. We can then calculate
the expected value of pairwise distances, $\langle d \rangle$, and the expected
square of pairwise distance, $\langle d^2 \rangle$, from which we will have the
combinatorial polarization 

\begin{equation}
  P_c = \langle d^2 \rangle - \langle d \rangle^2
\end{equation}

We will show that $P_c=\frac{1}{K}$ by showing that $\langle d \rangle = 1$ and
$\langle d^2 \rangle = \frac{K+1}{K}$. Before we do that, we will derive functions
to help us count the number of pairs separated by a particular distance, and
to calculate distances between vertices on subcubes $Q_L \subseteq Q_K$. First,
we denote the total number of pairwise distances as $n=\frac{N(N-1)}{2}$ where $N$ is
the number of agents. The number of $L$-subcubes $Q_L \subseteq Q_K$ is

\begin{equation}
  n_s(L, K) = 2^{K-L} {K \choose L}
  \label{eq:ns}
\end{equation}
\noindent
This results from the fact that at all $2^K$ vertices of $Q_K$, ${K \choose L}$
subcubes can be created by choosing $L$ nodes adjacent to the vertex. This 
gives us $2^K {K \choose L}$ subcubes. This overcounts since each 
generated subcube was generated once for each of its $2^L$ vertices. So we must
divide by a factor of $2^L$, giving us the expression in 
Equation~\ref{eq:ns}.

Within $Q_L$, the number of pairwise distances where agents occupy antipodal
vertices is 

\[
  n_a'(L, K) = 2^{L-1} \left(\frac{N}{2^K}\right)^2
\]
\noindent
There are $2^{L-1}$ pairs of antipodal vertices in $Q_L$. In the large
$N$ limit, agents are distributed in equal number to each vertex of $Q_K$.
Then, the number of agents in a single vertex is $\frac{N}{2^K}$, so the number
of pairwise distances between any two antipodal pairs is $\left(\frac{N}{2^K}\right)^2$.
The total number of antipodal pairs across all $Q_L$ is then

\begin{equation}
  n_a(L, K) = n_a'(L, K) n_s(L, K).
\end{equation}
\noindent
Finally, the distance between agent opinions $\vec s_1$ and
$\vec s_2$ in antipodal vertices of $Q_L$ is

\begin{equation}
  d_a(L, K) = \frac{1}{K} \sum_{k=1}^{K} |s_{1k} - s_{2k}| = \frac{2L}{K}
\end{equation}
\noindent
since any antipodal vertices of $Q_L$ share $K-L$ opinion coordinates, and the
maximum magnitude of difference on a single opinion dimension is 2. 

With these quantities we can write the expected value of pairwise distance,

\begin{equation}
  \langle d \rangle = \frac{1}{n} \sum_{L=1}^K n_a(L, K) d_a(L, K).
\end{equation}
\noindent
Simplifying and taking $N \rightarrow \infty$, this becomes
\[
  \langle d \rangle = \frac{(K-1)!}{2^{K-1}} \sum_{L=1}^K \frac{1}{(K-L)!(L-1)!}
\]
\noindent
Using the identity
\[
  \sum_{L=1}^K \frac{1}{(K-L)!(L-1)!} = \frac{2^{K-1}}{(K-1)!},
\]
\noindent
we find $\langle d \rangle = 1$. Calculating $\langle d^2 \rangle$ proceeds 
similarly, beginning with

\begin{equation}
  \langle d^2 \rangle = \frac{1}{n} \sum_{L=1}^K n_a(L, K) d_a(L, K)^2.
\end{equation}
\noindent
Simplifying and taking $N \rightarrow \infty$, this becomes
\[
  \langle d^2 \rangle = \frac{(K-1)!}{2^{K-2}K} \sum_{L=1}^K \frac{L}{(K-L)!(L-1)!}.
\]
\noindent
With the identity
\[
  \sum_{L=1}^K \frac{L}{(K-L)!(L-1)!} = \frac{2^{K-2}(K+1)}{(K-1)!}
\]
\noindent
we find $\langle d^2 \rangle = \frac{K+1}{K}$. So

\begin{equation}
  P_c = \langle d^2 \rangle - \langle d \rangle^2 = \frac{K+1}{K} - 1 = \frac{1}{K}.
\end{equation}

\section*{Data Availability}
Data used for our analyses is available for download ($\sim$14GB) from 
\url{http://mt.digital/static/data/polarization_v0.1-data.tar}. 

\section*{Conflicts of Interest}
There are no conflicts of interest for either author.

\section*{Acknowledgements}
Computational experiments were performed on the MERCED computing cluster,
which is supported by the National Science Foundation [Grant No. ACI-1429783].

\clearpage

\bibliographystyle{apacite}

\setlength{\bibleftmargin}{.125in} \setlength{\bibindent}{-\bibleftmargin}

\bibliography{ComplexitySpecialIssue.bib}

\end{document}